# Dirac's Spectrum from Newton's Laws in Graphene

Vardan Apinyan [*], Mane Sahakyan

Institute of Low Temperature and Structure Research, Polish Academy of Sciences, PO. Box 1410, 50-950 Wrocław 2, Poland; E-Mails: v.apinyan@intibs.pl; m.sahakyan@intibs.pl

* **Correspondence:** Vardan Apinyan; E-Mail: v.apinyan@intibs.pl

**Abstract**

The present report provides a phenomenological theory for monolayer graphene, in which two worlds, the quantum and the classical, meet and complete each other in the most natural manner. In the present report, electron mass-vortex representation is introduced, and the surface tension excitation states in the monolayer graphene are defined. The band-mass of the electrons at the Dirac's point was calculated through the mathematical introduction of a mass-dispersion relation. As a consequence, the Dirac's energy dispersion in monolayer graphene from classical Newton's law was obtained. Within the scope of the semiclassical theory, it was demonstrated that there was the presence of surface spin tension vectorial field, which, possibly, closely links the surface tension and the spin tension states of the helical surface. The surface tension associated with the confinement of the electron band mass-vortex at the Dirac's point was calculated, and the relaxation time of the surface tension state was predicted with accuracy. Moreover, it was demonstrated, phenomenologically, that the manifolds on $S^{(6)}$ are not integrable (which had been a long-standing problem in the group theory). The principal reason for this was attributed to the irreducibility of the spinorial group $Spin(6)^{R}$ at the Dirac's point, because of band-mass formation and confinement via the gravitational field.

## 1. Introduction

Monolayer graphene has attracted considerable attention in the last few years, because of its extraordinary physical properties [1]. Recent advances in the nanotechnological studies on graphene led to the fabrication of graphene-based adjustable metamaterial devices, such as biased graphene, graphene nanorings, graphene nanodisks, graphene heterostructures, graphene cross, and graphene sheet, which exhibit excellent optical, electrochemical, and thermal properties [2-10]. These properties of graphene are principally a result of its electron energy spectrum: at the $\boldsymbol{K}$-point, of its Brillouin Zone (BZ), the electron and hole bands touch one another, while in the vicinity of that point, they split linearly in the wave vector $\boldsymbol{k}$. Therefore, the spectrum appears similar to the spectrum of relativistic Dirac particles. The inequivalent Dirac's points $\boldsymbol{K}$ and $\boldsymbol{K}'$ in single-layer graphene are stable against perturbations which preserve the discrete spacetime inversion symmetry [11]. The degeneracy of these points is secured by the point-group symmetry of the hexagonal lattice. Recently, monolayer transition-metal dichalcogenides were proposed for the study of Dirac's physics, with excitonic optical addressability at specifiable momentum, energy, and pseudospin [12]. It was demonstrated that the valley pseudospin is coupled strongly to the exciton's center-of-mass motion through an effective electron-hole exchange mechanism. This coupling realizes a massless Dirac's cone with the chirality index $I = 2$ for the excitons inside the light cone, i.e. the bright excitons which realize a strong valley–orbital coupling in the order of magnitudes larger than the radiative recombination and momentum scattering rates. Furthermore, after performing the measurements for angular resolved photoluminescence (PL) in such a strong coupling regime, the pseudospin splitting from the valley–orbit coupling becomes spectrally resolvable. Recently, in the terahertz regime, a graphene-based metamaterial structure was developed as a wideband tunable reflectarray for generating orbital angular momentum vortex waves [13]. In particular, it was demonstrated that spiral phase distributions of the orbital angular momentum vortex waves, with values $l = \pm 1$, $l = \pm 2$, and $l = \pm 3$ for the corresponding modes, could be observed, and the intensity distributions were obtained in a wide range of frequency.

The inertial mass of vortices was discussed recently in [14], in terms of the energy of the unique kelvon quasiparticle in Bose and Fermi superfluids, and it was suggested that the inertial mass of a vortex originates due to the quasiparticles confined within the core of the individual vortex. With the consideration of the classical limit of large quantum numbers, a relationship between the Kelvin waves and the inertial mass of the classical vortices and vortex rings was obtained. Recently, it was reported that the non-local electrical response in graphene, driven by the chargeless modes, is sensitive to the quantities which are not directly accessible in the electrical transport measurements [15, 16]. Examples of these quantities are the spin currents and the valley currents. In particular, a giant non-locality close to Dirac point in graphene was observed in [17] while performing non-local magnetotransport measurements in the Hall bar geometry, and the observed large non-locality close to the Dirac point (which persists up to room temperatures) was attributed to the long-range flavor currents induced due to the lifting of the spin-valley degeneracy.

It is predicted that the strongly correlated electron systems obey the universal collision-dominated transport dynamics resembling that of the viscous fluids [18-21]. However, the study of such phenomena has been unsuccessful so far, due to the lack of known macroscopic signatures on electron viscosity [22-26]. In a recent report, vorticity was considered a signature of electron viscosity, which becomes a verifiable striking macroscopic dc transport behavior [27].

The evidence of the vortices which exhibit the electrons in graphene was provided experimentally [28] in a viscous flow of a Newtonian fluid inside a microfluidic device, with the obtaining of the existence of a theoretically-predicted threshold (see, for example, in [27]) at which the electron-vortices appear. The electronic transport behavior of these materials, with non-local current–field relation (negative resistance), was best described by low Reynolds number (Re) hydrodynamics, and the experiments in the low-Re limit of the Newtonian fluid enabled the observation of the unusual vortex flow dynamics which were unobservable experimentally in graphene or the other 2D electronic systems.

Another characteristic vortex-like shape of the bielectron states (a pair of two-dimensional (2D) mass-less Dirac–Weyl fermions) was reported in [29], in which a system of differential equations was derived for a novel type of quasiparticle– the bielectron-vortices, and it was demonstrated that the obtained system of equations had square-integrable solutions only at the Dirac's point, corresponding to the charged bound pairs with zero-energy.

Recently, a quantum mechanical treatment, reported in [30], suggested that the kinetic angular momentum of the electron-vortex beam is time-varying in a uniform magnetic field, which was in contrast to the classical solution for the rotationally-symmetric classical cyclotron motion of constant radius, and corresponding to kelvons, in 2D systems [31]. Nonetheless, the time-varying behavior of the kinetic angular momentum and the vortex-like motion of electrons along their classical cyclotron orbits could be explained within a classical model [32] constructed for the cyclotron motion under a uniform magnetic field with constant canonical angular momentum and inside a symmetric gauge.

In this report, a novel electron mass-vortex representation was introduced, in which electron is treated as a particle that possesses effective mass, rotation momentum $\boldsymbol{\Gamma} = 2\pi\, \boldsymbol{R} \times \boldsymbol{v}$ (see Landau-Lifshitz), and intrinsic mass-vortex with angular frequency $\Omega_{\Phi}$. *It was supposed that the circulation of the mass-vortex of the electron leads to the precession of its mechanical spin-momentum*. Therefore, for a single, free electronic state [33], the radius of spin precession is equal to the radius of its mass-vortex, while for the electrons in the solids, these radii could differ considerably, as would be described later in the example of monolayer graphene. It will be demonstrated how the gravitational vector field, acting in the mass-dispersion representation, leads to the enhancement of non-additive band-mass at the Dirac's point in the reciprocal space. Furthermore, to each point $\boldsymbol{k}$ in the reciprocal space $\mathbb{R}_{\boldsymbol{k}}^{(2)}$, a given surface area $ds$ in the real Cartesian space $\mathbb{R}_{\mathrm{C}}^{(2)}$ was associated. Such a construction, composed of two perpendicular spaces, allows the correct estimation of the localization radius of the electron close to the Dirac's point. It was determined to be much smaller compared to the radius of the electron itself, $r_s = \frac{e^2}{m_e c^2}$. Furthermore, by considering the surface tension excitation associated with the probe-electron at the vicinity of the band mass-vortex, at the Dirac's point, the Dirac's linear dispersion relation was obtained using the combination of Newton's gravitational law and the classical Hook's law, which ensured the stability and the possibility of a given excited state, close to the Dirac's point $\boldsymbol{K}$. This is a unique approach which provides Dirac's spectrum classically, without the requirement of addressing the tight-binding treatment of the problem, and one requires referring only to the concept of the electron's mass-vortex, which was postulated "ad hoc" by our research group. Here, it was demonstrated that such a representation of the electron provides a correct description of the surface tension states in monolayer graphene and a reasonable result for the excitonic decoupling time, at the Dirac's point. In addition, the fundamental mathematical results obtained following the theory presented here and concerning a series of unresolved problems in the spin-group representation were discussed briefly.

The logical construction of the present report is as follows: first, the mass-vortex representation of the electron is introduced, followed by discussing the mass-acceleration phenomena toward the Dirac's point in graphene and the demonstration of the formation of non-additive band-mass via gravitation field. Next, obtaining the Dirac's linear band dispersion relation in deep connection to Newton's classical mass attraction law, and the calculation of the surface tension for the electron mass-vortices and the decoupling time of the surface tension state, corresponding to the excitonic mass-vortex, is described. Finally, there is a discussion on the principal results of the theory presented here, concerning the spin-group representation. In the end, a brief conclusion to the report is provided.

## 2. The Mass-Vortex Representation of the Electron

It is well recognized that the electrons in monolayer graphene, owing to the crystal structure of the material they are located in, obey the quasi-relativistic law of movement. The pristine graphene monolayer, when alone, exhibits the perfect metallic electronic properties, while under applied electric field, it becomes semi-metallic with the associated small band overlaps of the order of 4–20 meV. The quasi-relativistic nature of the particles in graphene films is associated, first of all, with the zero real mass of the electrons at the $\boldsymbol{K}$ point in the Brillouin Zone. The reason for this is mathematically quite simple: the second derivative of the energy, culminating linearly at the $\boldsymbol{K}$ point, diverges smoothly at this point. Meanwhile, in the real crystal, the effective mass-tensor of the electrons acquires a minute, although finite value, owing to the crystal structure of graphene. The situation, however, is quite different in the bilayer graphene (BLG) structure, where there are four parabolic energy bands, two of which touch each other at the $\boldsymbol{K}$ point in the BZ. The second pair of the energy bands are separated by the minimal energy gap in the BLG, classifying the BLG system as an ideal semiconducting material with a minute bandgap (at times, BLG is also considered a normal metal with the largest minimal bandgap, characteristic of the normal metals). When the external electric field is applied to the BLG, the minimal gap becomes extremely large, while the zero-gap touching parabolae become separated, and the well-known "sombrero"-like shape appears in the band structure.

The general idea behind the proposed theory is to formulate a physical quantity which could be expressed using the parameters associated with the intrinsic properties of the mass-vortex (which governs the spin rotation of the electron) and the rotational momentum of the electron. It appears that such a parameter could be expressed as a product of surface tension $\Delta\epsilon$ [34] (in the usual terminology, this would be graphene's band energy per unit area in the 2D reciprocal space) and the interval of time $\Delta t$ during which the created surface tension excitation appears. The only prerequisite for the surface tension states to appear is the existence of a non-dispersive interface [35]. The creation of an interface leads to an increase in the local probability distribution (which describes the possible location of the particles within the system) at the boundary. Recently, it was demonstrated theoretically [36] that truncation of the gravitational forces and fields leads to the surface tension states (for this, the authors introduced a mechanical model of spacetime), regardless of the material composition and phase. Here, *the surface tension $\Delta\epsilon$ is nothing but energy, corresponding to the unit surface area, and created by the rotation of the mass-vortex*. The term "tension" is used here with the assumption that $q_e/m_e \equiv 1$, where $q_e$ denotes the charge of the electron. Indeed, the mass-vortex around the electron is generally charged and supports a minute, although finite, electric-field tension, which should also be included in the theory; however, this was out of the scope of the present treatment. In the problem considered here, it was supposed that the electron is neutral. Using a simple dimensional analysis, one would be convinced that the product $\Phi_E = \Delta\epsilon\Delta t$ could be expressed using the parameters that describe the intrinsic properties of the electron as a neutral particle. The following quantum-mechanical relation was defined:

$$\Phi_{\mathrm{E}} = \Delta\epsilon\Delta t = 4\pi|\mathbf{S}|\frac{\Omega_\Phi}{|\Gamma|}, \tag{1}$$

where $|\mathbf{S}|$ is the modulus of mechanical spin momentum, $\Omega_\Phi$ denotes the angular frequency of the free electron's mass-vortex, and $|\mathbf{\Gamma}|$ is the modulus of the electron's rotation momentum. Indeed, the dimension on the right-hand side of this expression is of the order of $[\Phi_E] = eV \cdot sec/m^2$, in SI units. If the time interval $\Delta t$ on the left-hand side of the Eq. (1) is extremely small, i.e., the case of instantaneous excitations, then the derivative of the quantity $\Phi_{\mathrm{E}}$ with respect to time provides the surface tension states that are directly associated with the band structure in graphene. The quantity $\Omega_\Phi$ is material-specific as it may be expressed as $\Omega_\Phi = \sqrt{\frac{\kappa_\Phi}{m_\Phi}}$, where $\kappa_\Phi$ is referred to as the vortex viscosity. In other words, parameter $\kappa_\Phi$ is just the spring constant of the given material where the vortex is placed in, and in the present case, it would be the spring constant of monolayer graphene $\kappa_G$. The quantity $m_\Phi$ denotes the mass of the electron's mass-vortex. Here, it was postulated that *the electron mass-vortex is separated at a certain distance $\Delta x$ from the center-of-mass of the electron*. Therefore, $\Delta x$ denoted the polarization of the mass-vortex. The mass-vortex representation of the electron, proposed in the present report, is presented in Figure 1, where the electron's mass-vortex is depicted in the form of a filled grey circle attached to the electron.

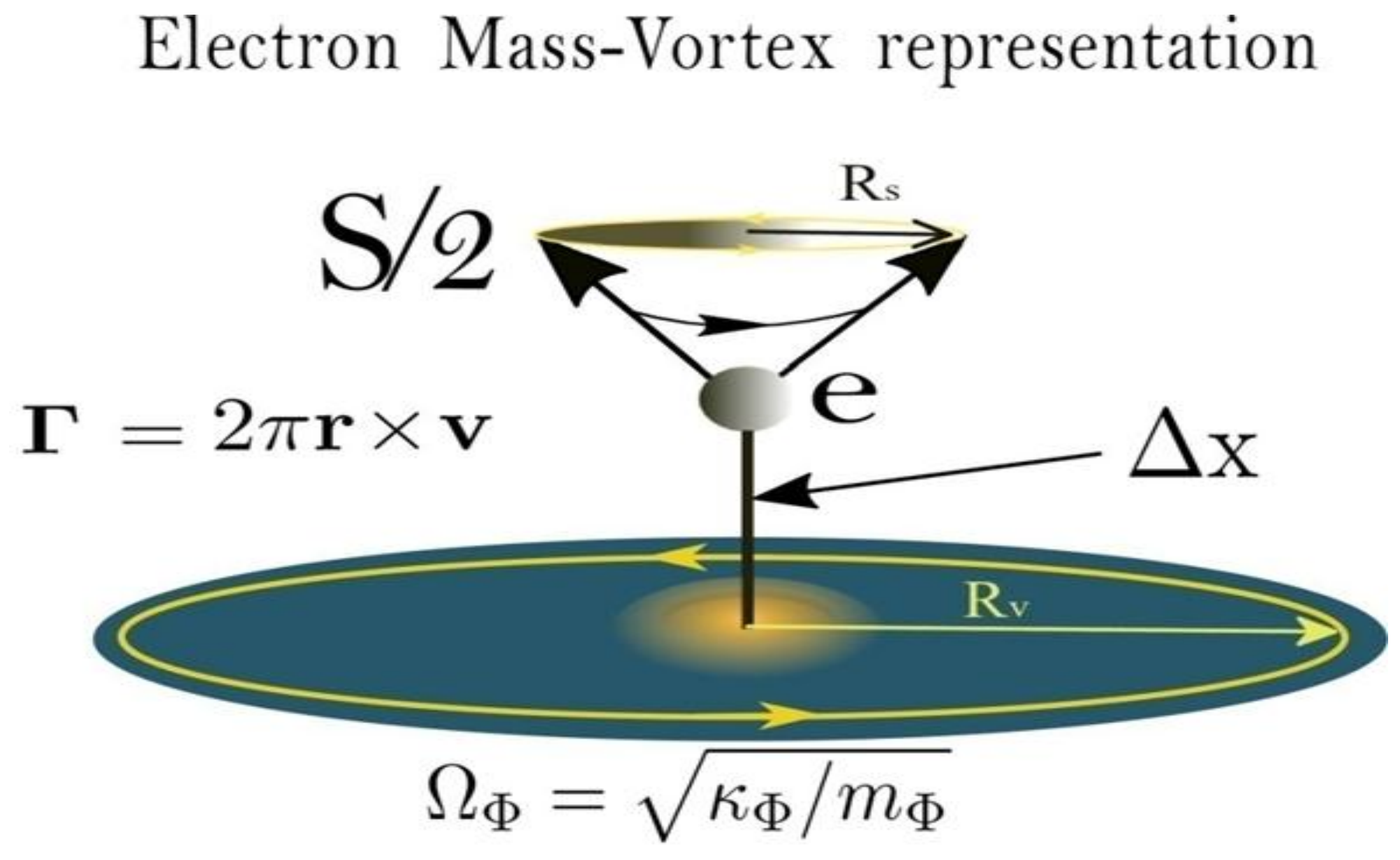

**Figure 1** The schematic representation of the electron's mass-vortex, with angular frequency $\Omega_\Phi$, leading to the rotational precession of the mechanical spin momentum $S = \hbar/2$, both of which are the intrinsic mechanical properties of the electron. The rotational momentum of the electron is expressed as $\mathbf{\Gamma} = 2\pi\, \mathbf{R} \times \mathbf{v}$.

It would be described later that the radius $R_v$ of the electron's mass-vortex could be considerably different from the electron's spin-precession radius $R_s$, and depends, in general, on the position of the reciprocal wave vector in the 2D $\boldsymbol{k}$-space. Figure 1 presents the general case where $R_v \neq R_s$ (i.e., the formed band-mass is different from the free electron's effective mass). In the case of the individual free-electron, $R_v{\sim}R_s$ . The physical quantity $\Omega_\Phi$ represents the angular frequency of the electron mass-vortex, which leads to the precession of the intrinsic mechanical momentum of the electron ($S$=$\hbar/2$) along the $z$-axis. The second parameter in Eq. (1) is the electron's rotation momentum $\mathbf{\Gamma}$, which is the electron's angular momentum $\mathbf{L}$ renormalized to the mass of the free electron $m_e^*$**,** i.e. $\mathbf{\Gamma} = \mathbf{L}/2\pi m_e^*$. It was also postulated that for each point $\boldsymbol{k}$ in the reciprocal space, the electron is localized in a certain region in real space where the circulation of the mass-vortex occurs. This localization region belongs to the 2D space $\mathbb{R}_\mathrm{C}^{(2)}$, which is the subject of the real Cartesian space, perpendicular to the reciprocal 2D $\boldsymbol{k}$ -space $\mathbb{R}_{\boldsymbol{k}}^{(2)}$. On the other hand, the 2D $\boldsymbol{k}$ -space is the subject of one-dimensional (1D)-sphere $S_{\boldsymbol{k}}^{(1)}$ in the reciprocal space.

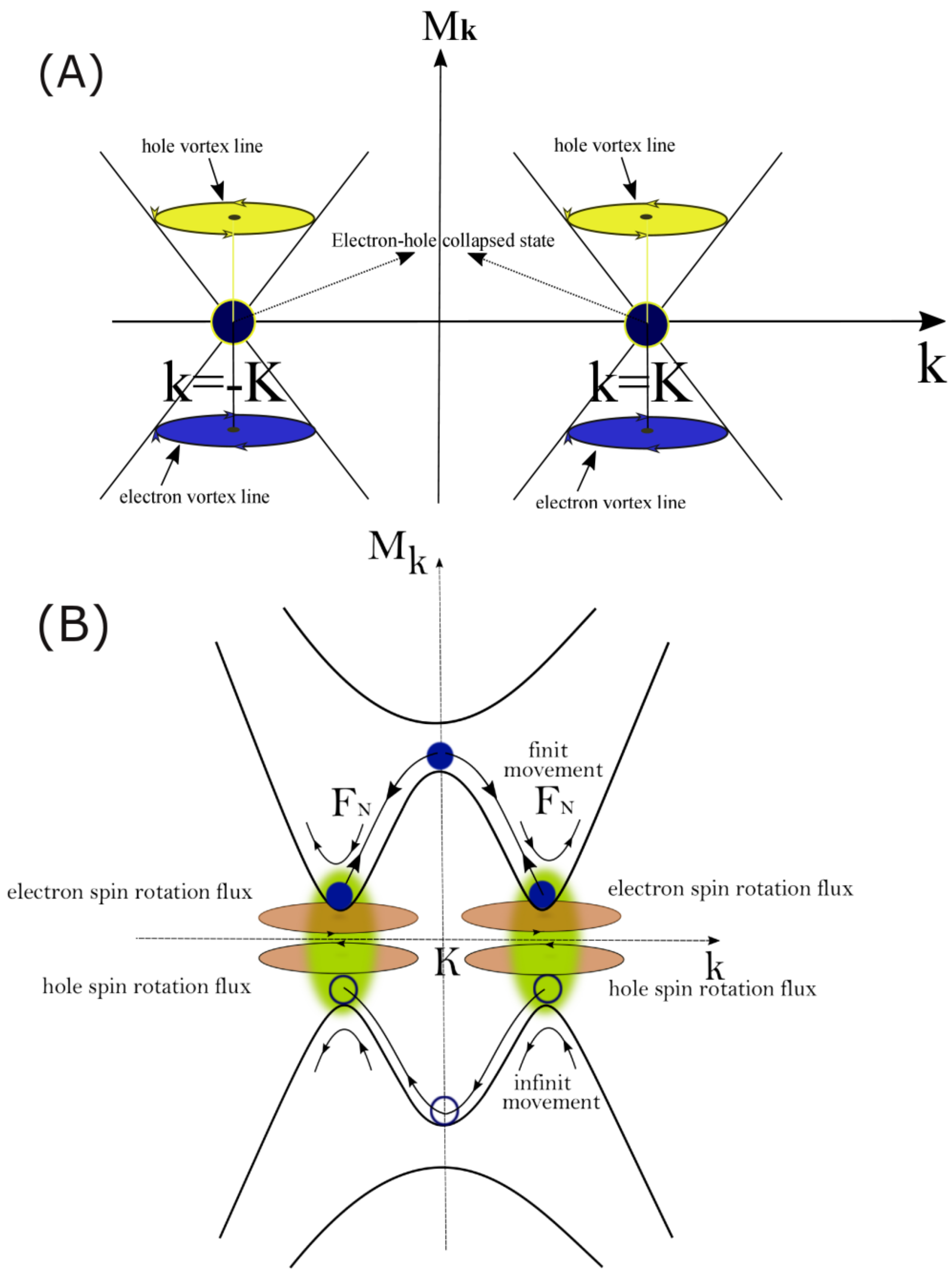

**Figure 2** The dispersion bands in monolayer graphene (at the top) and the biased BLG (at the bottom). The electron and the hole mass-vortices are depicted in the picture, and the collapsed excitonic states are depicted in the form of blue balls, at the Dirac's points, in the upper picture. The picture at the bottom depicts the excitonic formations between the conduction electrons and the valence band holes in the BLG. The positions of the mass-vortices are in good agreement with the semi-metallic nature of monolayer graphene.

An interesting description of the surface tension excitation states is provided in [37], where the author applies the concepts of physical chemistry of surfaces and surface tension to spacetime. It was suggested that the spacetime surface tension description could serve as the basic mechanical model for several phenomena, with the time-like motion provided by the Klein-Gordon type equation and the space-like motion having the form of Schrodinger-like equation. In addition, it was suggested that the Bohr-Heisenberg-type uncertainty relation for a change in the surface tension and a change in strain could be embedded within the classical Hooke's law for the mechanical medium of spacetime (refer to the description in [38]).

Figure 2 presents the mass-vortices for the electrons and holes at the non-equivalent Dirac's points $\boldsymbol{K}$ and $-\boldsymbol{K}$ in the unbiased monolayer graphene, with zero bandgap (the upper picture (A) in Figure 2) and in the case of biased BLG (the lower picture (B) in Figure 2). Indeed, the electron's mass-vortex is situated in the valence band in the band structure of monolayer graphene, while the hole mass-vortex is situated in the conduction band, with both the vortices being equidistant from the corresponding particles (refer to the upper picture (A) in Figure 2). Therefore, the semi-metallic nature of monolayer graphene is well-described in the mass-vortex representation, such that it was postulated that the mass-vortices are the inseparable counterparts of the electrons. The upper picture in Figure 2 depicts the collapsed excitonic state in the form of blue balls at the Dirac's points $\boldsymbol{K}$ and -$\boldsymbol{K}$, while in the lower picture, the excitonic correlations are apparent and the excitons are depicts in the form of light-green clouds formed between the conduction and the valence bands of the gapped band structure. Moreover, in the case of BLG, the electron and hole mass-vortices are located in the gap region, and their mutual positions and binding strengths define the stability of the tunable gap value. Therefore*, the band gap in BLG is formed from the mass-vortex fields created by the electrons and holes.* Furthermore, the radius $R_v$ of the electron's mass-vortex circulation was estimated. In case of monolayer graphene, the impulsion of the electron at the $\boldsymbol{K}$-point in the BZ is not well defined, as the first derivative of the energy dispersion $\varepsilon(\boldsymbol{k})$ with respect to the wave vector does not exist at the Dirac's point. Therefore, the electron, excited from the valence band, could continue traveling along one of the two branches in the conduction band, which form the helical surface appearing to arise from the linear band structure in monolayer graphene. Consequently, the tangential (or covariant) component of the velocity vector of the electron in the band structure would be pointed along the positive direction of the $\boldsymbol{k}$ wave vector axis, or in the opposite direction.

The left-hand side of Eq. (1) describes the decaying surface tension states with a decay time of the order of $\Delta t$, while the right-hand side is a function of the properties of the electron such as the modulus of spin momentum, the rotation momentum vector $\mathbf{\Gamma}$ (otherwise, the orbital momentum in the units of the mass of the electron), and the electron's mass-vortex angular frequency $\Omega_\Phi$ associated with the circulation of the band mass-vortex of the electron. A similar relation, as in Eq. (1), could be written for the hole particles as well, with the corresponding parameters associated with the holes. It should be noted that the directions of rotation of the spin and the circulation of the mass-vortex of the *hole* particles are opposite to those of the *electrons*. Therefore, the product $\Phi_\mathrm{E}$ is a function of the internal parameters associated with the intrinsic construction of the electron (or hole). The time derivation of $\Phi_\mathrm{E}$ provides a quantity proportional to the surface tension, or the energy per unit surface area. Such a derivation in time is permitted only when the time-grid interval $dt_i$ is extremely small in comparison to the decay-time duration of the excitation, i.e. $\mathrm{dt_i} \ll \Delta\mathrm{t}$. This manifests physically as the variation of surface tension states on a time-scale much shorter compared to the attenuation time of the excited state. Eq. (1) is a composite analytical formula, which relates the surface tension states with the intrinsic parameters of the electron, and consequently, one could expect to have a relation among the mass of the electron mass-vortex, velocity of the electron, radius of the electron mass-vortex, and the spring constant $\kappa_\mathrm{G}$ of the mass-vortex in graphene. These results may be generalized for the case of BLG, and, in general, for the case of $\zeta$-layer graphene. The usual energy dispersion curves $\varepsilon(\boldsymbol{k})$ could be replaced with surface tension dispersion $\epsilon(\boldsymbol{k})$ (the vector $\boldsymbol{k}$ has two components $k_x$ and $k_y$). Indeed, those representations are equivalent, as the surface tension states are provided by $\epsilon(\boldsymbol{k}) = \varepsilon(\boldsymbol{k})/S_{\boldsymbol{k}}$ , where $S_{\boldsymbol{k}}$ represents the total surface area in the 2D reciprocal $\boldsymbol{k}$-space.

Let's consider another representation associated with reciprocal space in graphene, often referred to as "mass-representation" in the group theory [39]. Let's suppose a mathematical model of the dispersion curves associated with the masses of the electrons at each point on the linear dispersion lines. Therefore, the $M(\boldsymbol{k})$-dispersion is considered and the electron is abstracted as a particle without electrical charge, with the mechanical spin momentum and the given mass (the effective mass of the electron) moving with a given velocity $\boldsymbol{v_e}$ along the upper branches of the energy dispersion lines in monolayer graphene. It is clear that each electron, being a classical object, obeys the gravitation Galilean forces, which accelerate the electrons with the well-known universal acceleration constant $g$ ($g$ = 9.806 m/$s^2$). The gravitation forces are acting on each of the electrons, in the direction perpendicular to the $\boldsymbol{k}$-axis and also perpendicular to the real 2D plane $(x, y)$. Let's suppose the existence of two 2D spaces which are perpendicular to each other. One of them is the reciprocal space $\boldsymbol{k} = (k_x, k_y)$ [which is the subject of the $S_{\boldsymbol{k}}^{(1)}$ sphere], and the other one is the real Cartesian space $\boldsymbol{r} = (x, y)$ [which forms the hypersurface in $S_C^{(1)}$]. These spaces are naturally perpendicular because of the construction of the reciprocal space in solid state physics. Therefore, there are, in total, three types of representations of the band structure of the electrons– the usual energy band structure $\varepsilon(\boldsymbol{k})$-dispersion, the surface tension representation $\epsilon(\boldsymbol{k})$, and the mass-dispersion representation $M(\boldsymbol{k})$. In each of these cases, one has to deal with the 2D vectorial space $\boldsymbol{k}$ and one of the scalar fields [$\varepsilon(\boldsymbol{k})$, $\epsilon(\boldsymbol{k})$ or $M(\boldsymbol{k})$]. When the normal vectors $\boldsymbol{n}_\varepsilon$, $\boldsymbol{n}_\epsilon$, and $\boldsymbol{n}_M$ are added to the scalar fields stated above, one obtains a quasi 3D or 2+1-dimensional vector fields for each of the discussed cases. Hereafter, the present report will deal mainly with the states in the surface tension representation $\epsilon(\boldsymbol{k})$.

## 2. Mass-Acceleration toward Dirac's Point

### *2.1 The Non-linear Dependence of Mass on the Number of Electrons*

Figure 3 depicts the helical surfaces for the electrons and holes in monolayer graphene and under zero external fields. The 2D $\boldsymbol{k}$-space, which forms the hypersurface in 1-sphere $S_{\boldsymbol{k}}^{(1)}$, is illustrated in the expanded form [the plane $(k_x, k_y)$ in Figure 3] in the BZ in graphene. The infinitesimal surface $dS_{\boldsymbol{k}}$ is depicted in the 2D plane of the wave vector $\boldsymbol{k}$, and represents the electron's localization size in the $\boldsymbol{k}$-space. In the same image, the infinitesimal surface $dS_l$ is presented, which defines the electron's localization size in the real space. Using the reciprocal and real Cartesian spaces, in which the electrons and holes are moving simultaneously, one is able to construct a direct product of these spaces. The resulting space $\mathbb{R}^{(4)}$, which is the subject of the 3-sphere, is a direct product of two orthogonal spaces $\mathbb{R}_\mathrm{C}^{(2)}$ (2D Cartesian space) and $\mathbb{R}_{\boldsymbol{k}}^{(2)}$ (2D reciprocal space): $\mathbb{R}^{(4)} = \mathbb{R}_\mathrm{C}^{(2)} \times \mathbb{R}_{\boldsymbol{k}}^{(2)}$. As would be demonstrated here, this space is indeed a 1D hypersurface, i.e. the 0-sphere, of the dimension of length, in the real Cartesian space $C$. Here, the band-mass at the Dirac's point in the mass-dispersion approximation of the band structure in graphene would be calculated and would be demonstrated as a non-linear function of the total number of particles in the conduction band. When the electrons are considered neutral particles with the given spin, they obey the gravitational forces which act on each of the $N$ electrons (it is supposed that two branches in the linear band structure in graphene have an equal number of electrons) in the conduction band. Therefore, it is postulated that the electrons in graphene obey the gravitational force as do all the particles in nature that possess mass. Recently, demonstration of the emergent gravitational field was reported in monolayer graphene [40, 41] and in Dirac/Weyl semi-metals [42], and the authors of these reports confirmed that the geometry experienced by the fermionic quasiparticles in graphene is related to the varying 2D Weitzenbock geometry [43] expressed through elastic deformation. This implies that there exists an emergent teleparallel gravity in graphene. The electrons, as the usual classical particles, are accelerated under the gravitational field with acceleration $\boldsymbol{g}_{\perp}$ which is a component of the vector of gravitational acceleration $\boldsymbol{g}$ along the mass-dispersion branches, as presented in Figure 4. Hypothetically, the mass-dispersion branches are considered the possible medium (the glue) through which the gravitational forces contribute and cumulate. The projection of each of the gravitational acceleration vectors along the dispersion branches leads to the total band-mass acceleration toward the Dirac's symmetry point $\boldsymbol{K}$ in the BZ, as illustrated in Figure 4. The center-of-mass for each pair of equipotential electrons, on different symmetry branches, in the $M(\mathbf{k})$ −dispersion image depicted in Figure 4, is centered on the line perpendicular to the $\boldsymbol{k}$-axis in the reciprocal space, which corresponds to the perpendicular real-space axis $|\boldsymbol{k}|ds$ in the hypothetical space $\mathbb{R}^{(4)}$ of the dimension of length, i.e. $\dim[\mathbb{R}^{(4)}] = cm$.

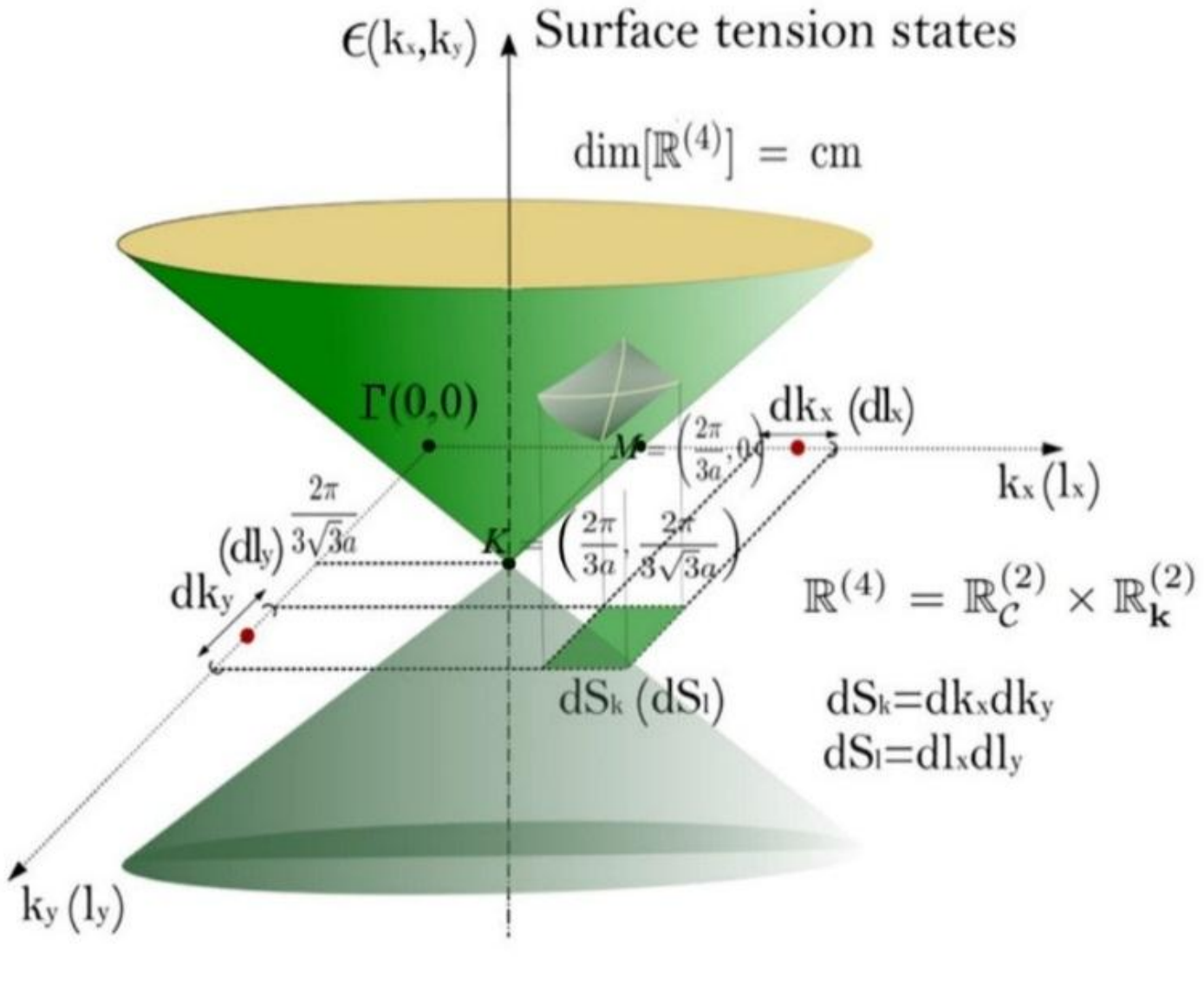

**Figure 3** The helical surface of conduction band electrons (upper conical surface) and valence band holes (lower conical surface), in the surface tension representation of the electronic band structure in single-layer graphene. Positions of the high-symmetry points are in 2D $\boldsymbol{k}$-space, while the direct product of the two parallelly-existing spaces is presented in the picture. The resulting space has the dimension of distance, i.e. $\dim[\mathbb{R}^{(4)}] = cm$ .

Space $\mathbb{R}^{(4)}$ is the subject of 3-sphere and a direct product of two orthogonal spaces as stated earlier. Let's begin to discuss the effective band-mass of the electrons moving in the conduction band of graphene. In order to maintain clarity in the discussions, the electrons in one of the upper branches of graphene dispersion, let's say the left linear dispersion line in Figure 4, were considered, supposing an odd number of total electrons, i.e., $N = 2k + 1$. In this manner, one obtains the equal number of electrons $N' = [N/2] = k$ (here, the parenthesis $[\dots]$ implies the integer part of the number $N/2$) in each branch in the upper part of the band structure, and one electron, fixed at the Dirac's point $\boldsymbol{K}$. Next, the electron localized at the Fermi surface of graphene (Figure 5), which is the last electron when counting from the Dirac's point and the closest neighbor electron with the number $[N/2] - 1$ on the same branch of the dispersion line, is considered. In order to maintain simplicity, the integer numbers were attached to the electrons beginning from the Dirac's point $\boldsymbol{K}$, and the number 0 was ascribed to the electron situated at $\boldsymbol{K}$. This enumeration procedure is illustrated in Figure 5. As stated earlier, the usual vectorial-addition rules were supposed for the gravitational field (acting through the mass-dispersion branches), as in the case of several corps in classical mechanics put together on the same rigid platform. The sum of the vectors $g_\perp$ at the position of the second electron, with the number $[N/2] - 1$, is $2g_\perp$. The total gravitational force acting on that electron toward the Dirac's point could be expressed as $F_2 = 2m^* g_\perp = m_2 g_\perp$, where $m_2 = 2m^*$. This effect of the gravitational field leads to an equivalent assumption to have a quasiparticle at the position $[N/2] - 1$, with a mass twice as large as the effective mass $m^*$ of the electron at the Fermi level. When the next electron is considered, one obtains $F_3 = m_3 g_\perp$, where $m_3 = 3m^*$, and so on. When this procedure is continued for the whole left branch in the band structure, up to the Dirac's point $\boldsymbol{K}$, one obtains the following expression:

$$\begin{aligned} \mathrm{F}_{01} &= (m_{\left[\frac{N}{2}\right]} + m_{\left[\frac{N}{2}\right]} + m_{\left[\frac{N}{2}\right]-1} + m_{\left[\frac{N}{2}\right]} + m_{\left[\frac{N}{2}\right]-1} + m_{\left[\frac{N}{2}\right]-2} + \cdots \\ &\dots + m_{\left[\frac{N}{2}\right]} + m_{\left[\frac{N}{2}\right]-1} + \cdots + m_1 + \mathrm{m}_0^{\mathrm{D}}) = m^* g_\perp (1 + 2 + 3 + \cdots + \left[\frac{N}{2}\right] + 1) \\ &= m^* g_\perp \frac{\left(\left[\frac{N}{2}\right]+1\right)\left(\left[\frac{N}{2}\right]+2\right)}{2} \quad . \end{aligned} \tag{2}$$

As an example, if there are a total of nine $[N = 9]$ conduction electrons, then $\mathrm{F}_{01} = 15m^* g_\perp$. Therefore, it is observed that the effect of the gravitational force leads to a large mass at the Dirac's point $\boldsymbol{K}$ which is not equal to $5m^* g_\perp$ as was expected when simply adding the masses of the electrons in the left-branch, rather the mass is three times larger. Eq. (2) also reveals that the total mass of the electrons at the Dirac's point arising from the left part of the linear band structure is $M_{Band} = m^* \frac{\left(\left[\frac{N}{2}\right]+1\right)\left(\left[\frac{N}{2}\right]+2\right)}{2}$, and it is a non-linear function of the total number of electrons $N$. The effect of the growth of the mass of the electrons due to the gravitational force is referred to as the band-mass acceleration toward Dirac's point $\boldsymbol{K}$. In order to calculate the total band-mass at the Dirac's point, one should also consider the total gravitational force $\mathrm{F}_{02}$ arising from the right branch. Its contribution is the same as presented in Eq. (2) (this is, of course, true for the non-deformed and pure monolayer graphene, without impurity atoms and defects). Therefore, the total force acting at the $\boldsymbol{K}$-point in the BZ, is expressed as $F = \sqrt{\mathrm{F}_{01}^2 + \mathrm{F}_{02}^2 - 2\mathrm{F}_{01}\mathrm{F}_{02}\cos\vartheta} = M_{Band} g_\perp$. Here, the relationship among the angles has been used: $\vartheta = \pi - 2\alpha$, where $\alpha$ represents the angle between the center-of-mass axis (Figure 4) and the mass-band (energy band) on the conical surface.

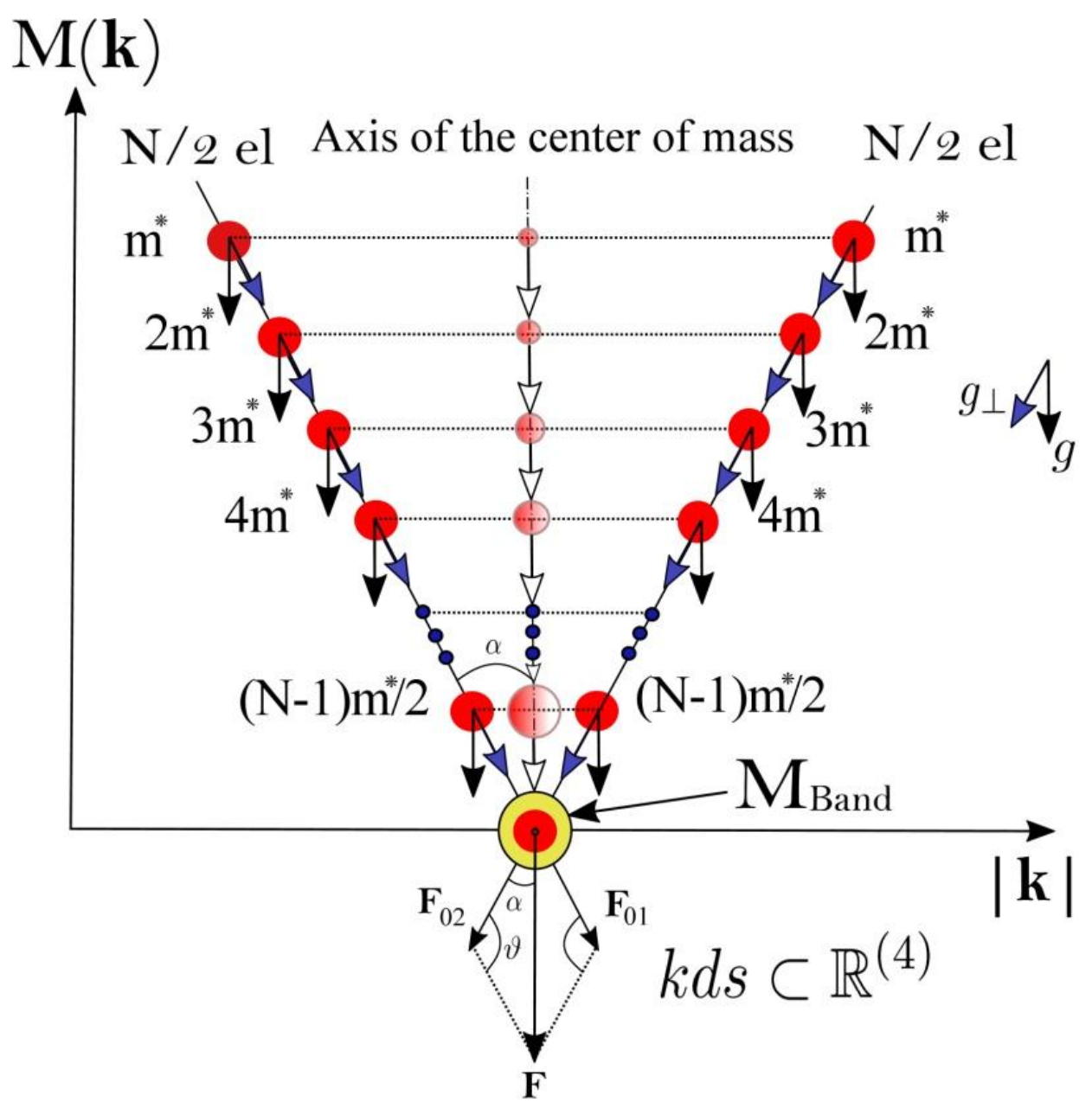

**Figure 4** Center-of-mass description in monolayer graphene. The effective mass-dispersion representation $M(\mathbf{k})$ is presented in the image; the red points represent the neutral electrons. Each branch in the upper band in graphene has $[N/2]$ electronic states. The total number of electrons is $N$, and it is supposed to be an odd number. The vertical holly arrows depicted in the middle of the band structure indicate the growing band-mass toward the Dirac's point $\mathbf{K}$.

Furthermore, the total force is expressed as $F = 2|F_0 \cos\alpha|$ [here, $F_0 = F_{01} = F_{02}$, because of the symmetry of the energy bands in non-deformed graphene]. The factor $|\cos\alpha|$ may be expressed in terms of the physical parameters in the system. Indeed, from the condition to have a classical particle at the Fermi level, it may be estimated that $\cos\alpha = m^* v_F^2 / 2\hbar\, v_F k_F = 0.5$ (Figure 4), where $v_F$ denotes the Fermi velocity of the electron with mass $m^*$ which is localized at the Fermi level with position number $[N/2]$ (see the outermost position of the electron in Figure 5). Next, using the expression for force $F_0$ (the total gravitational force that acts along the left dispersion branch) provided in Eq. (2), one obtains $F = M_{Band} g_\perp$, where $M_{Band}$ represents the band-mass at the Dirac's point and is expressed as follows:

$$M_{Band} = m^* \frac{\left(\left[\frac{N}{2}\right]+1\right)\left(\left[\frac{N}{2}\right]+2\right)}{2} \tag{3}$$

Here, $m^*$ is the effective mass of the electron in monolayer graphene [44]. Analogically, band-mass may be associated with any given point in the $\boldsymbol{k}$-space. For instance, in Figure 4, the band-mass at the position of the electron with number $[N/2] - 2$ (third electron from the Fermi level) is $M_{Band}^{[N/2]-2} = 6m^*$. Using the same methodology, it is possible to write the relation [1] for the real space coordinates (i.e., $|\boldsymbol{k}|ds$) of the center-of-mass of all the electrons in the BZ, which would be as follows:

$$R_C = KdS_D = \frac{\sum_{i=1}^{N} M_i^* k_i ds_i}{\sum_{i=1}^{N} M_i^*} \tag{4}$$

where the band-masses $M_i^*$ are strongly dependent on the position index $i$. In this context, there is a violation of the usual mass-addition rule (which states that the sum is a linear function of the number of pieces contributing to the sum) which is widely accepted in the classical physics, even though the gravitational vector-field additivity is supposed to be fulfilled in this case.

Next, for the first treatment, one may suppose that $ds_i \equiv ds$ for all the electrons in the given upper branch of the band structure. In the most general case, it does not hold true, because with the growth of the band-mass of the electrons toward the Dirac's point, the size of their localization on the real line $kds$ would also grow. An important point here is the denominator in Eq. (4), which provides the total mass of the electrons in monolayer graphene, which, as has been realized, is not additive in this case. The symmetry axis is passing through the Dirac's points, and the total center-of-mass in the upper band in monolayer graphene is situated on the axis $|\boldsymbol{k}| = K$. This is well-illustrated in Figure 4 and 5. Indeed, the gravitational field acts in the real space because, as discussed earlier, the movement of the electron in the $\boldsymbol{k}$ -space is accompanied by the simultaneous parallel movement along the real axis $|\,\boldsymbol{k}\,|ds$ in the four-dimensional (4D) space $\mathbb{R}^{(4)}$ which has the dimension of length.

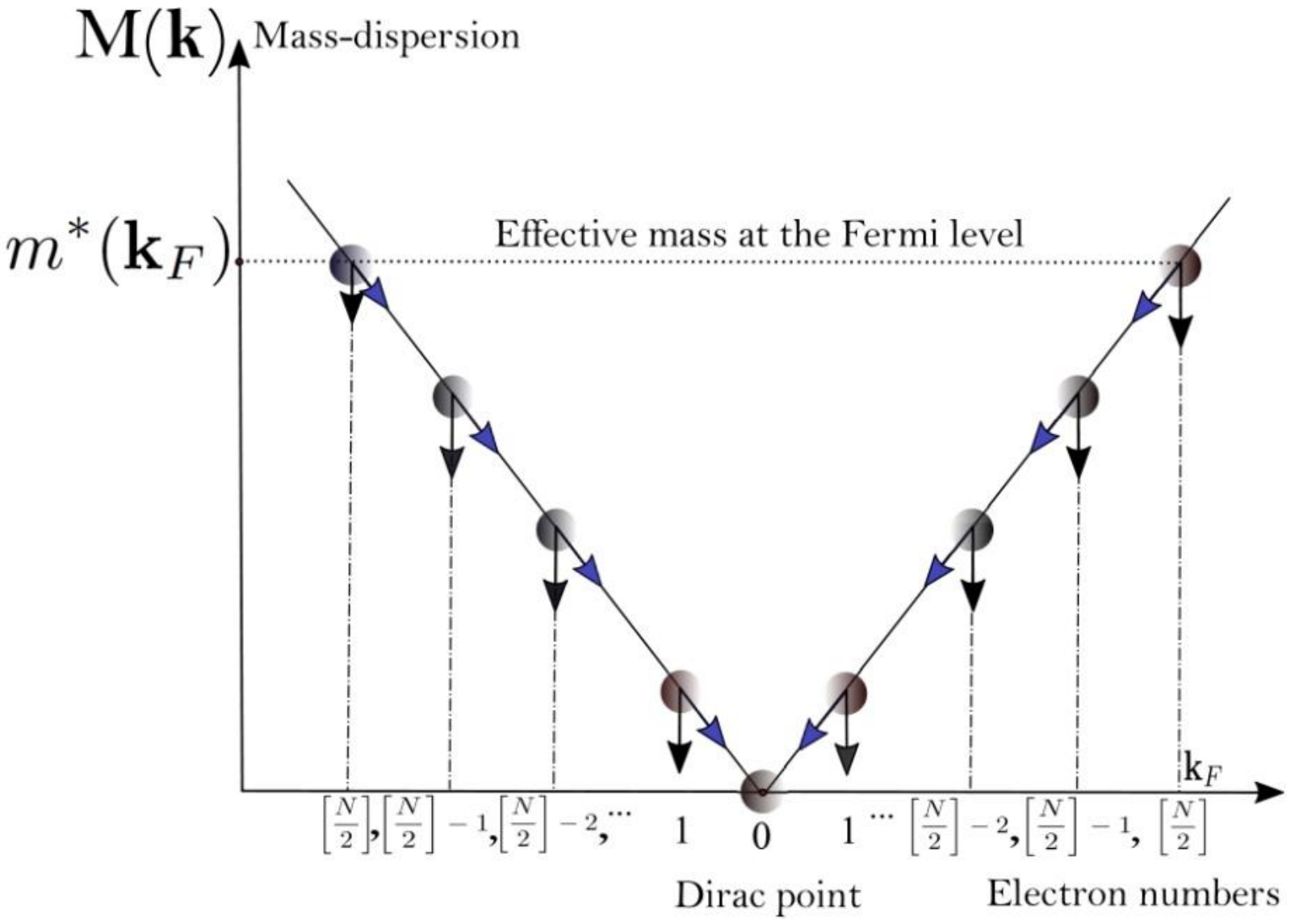

**Figure 5** Electron enumeration in the mass-dispersion scenario in monolayer graphene. The electrons are enumerated, beginning from Dirac's point $\boldsymbol{K}$. The last electrons in the enumeration are the electrons situated precisely on the Fermi level. The black arrows indicate the direction of the gravitational force acting on the electrons, while the blue arrows indicate the projection of the gravitational force along the dispersion axis.

As stated earlier, the procedure of calculation for the band-mass could be generalized for the case of $\zeta$-layer graphene heterostructures. Then, the band-mass for the $\zeta$ -layer graphene is as follows:

$$M_{Band}^{\zeta} = \sum_{i=1}^{\zeta} \sum_{j=1}^{\left[\frac{N}{2}\right]+1} M_{ij} = \zeta m^* \frac{\left(\left[\frac{N}{2}\right]+1\right)\left(\left[\frac{N}{2}\right]+2\right)}{2} \tag{5}$$

where the first index $i$ in the mass-tensor components $M_{ij}$ describes the layer in multilayer graphene, while the second index $j$ denotes the number of electrons in the single layer of the multilayer heterostructure. The coefficients $M_{ij}$ form a mass-tensor of the dimension $\zeta \times (\left[\frac{N}{2}\right] +$ 1). For instance, in the case of BLG, $M_{BLG} = m^* \left(\left[\frac{N}{2}\right] + 1\right)\left(\left[\frac{N}{2}\right] + 2\right)$, as the number of layers $\zeta$ is equal to 2. It may be observed here that the number of layers (after neglecting the interlayer coupling and interactions) exerts no spectacular impact on the value of band-mass. The surprising non-additive nature of band-mass at the Dirac's point exerts strong effects on the whole physics associated with the electron mass-vortices in graphene and layered graphene heterostructures. The mass-acceleration ratio $\delta$, in the monolayer graphene, relative to the usual additive mass $M_{add}$ is expressed as:

$$\delta_\infty = \lim_{N\to\infty} \frac{M_{Band} - M_{add}}{M_{add}} \approx \lim_{N\to\infty} \left(\frac{N}{2} - 1\right) = N \to \infty \tag{6}$$

where, $M_{add}$ denotes the additive mass of the material in general; in the present case, $M_{add} = Nm^*$. On the contrary, the ratio $\delta$ is of the order of $\sim 1$, if one normalizes the absolute mass $M_{Band} - M_{add}$ to the band-mass $M_{Band}$, i.e.

$$\delta_f = \lim_{N\to\infty} \frac{M_{Band} - M_{add}}{M_{Band}} \approx \lim_{N\to\infty} \left(1 - \frac{2}{N}\right) = 1. \tag{7}$$

The coefficient which provides the ratio between the effective total number of electrons participating in the formation of the band-mass and the total number of electrons forming the additive mass is as follows:

$$\rho_{eff} = \frac{N_{eff}}{N_e} = \frac{\left(\left[\frac{N}{2}\right]+1\right)\left(\left[\frac{N}{2}\right]+2\right)}{2N} \ . \tag{8}$$

In other words, the coefficient $\rho_{eff}$ is the relative excess of the electrons, which principally lead to the formation of the band-mass at the Dirac's point $\boldsymbol{K}$. According to Eq. (8), the coefficient $\rho_{eff}$ is proportional to the total number of electrons in the BZ in graphene, i.e., $\rho_{eff} \sim N$.

### *2.2 Quasi-Spiral Movement of the Spin: Bundles of Spin*

A previous study [45] experimentally demonstrated that the in-plane spin-texture (Figure 6) emerges in graphene's Fermi surface at $\boldsymbol{K}$ and $\boldsymbol{K'}$ when aligning the bandgap to the Fermi level. Since this spin arrangement is a valley effect [46], it implies the cancellation of Berry curvatures at the points $\boldsymbol{K}$ and $\boldsymbol{K'}$. In turn, the gapped Dirac's cone reveals a hedgehog out-of-plane reorientation of the electron spins with opposite spin orientations in the upper and lower Dirac's cones [47].

Inspired by the above-stated experimental findings, the text ahead will demonstrate how the movement of the electron in the conduction band of monolayer graphene could be accompanied by spin bundle effects on the helical surface. *In general, there is a difference between the movement of the electron and its spin on the helical surface*– the electrons are moving along the left and right branches which form the two opposite sides of the helical hypersurface in $S^{(1)}$, while the respective spins of these electrons are rotating along the lines on the helical surface. Meanwhile, when the electron is moving from the point $\boldsymbol{k}$ to the point $\boldsymbol{k} - \boldsymbol{dk}$ (here, $\boldsymbol{dk}$ is an infinitesimal increment) on the $|\boldsymbol{k}|$-axis, its spin could rotate along the corresponding path on the helical surface and return at the position of the electron at point $\boldsymbol{k} - \boldsymbol{dk}$, acquiring a phase equal to $2\pi$. At the initial state position at point $\boldsymbol{k}$**,** the spin of the electron is oriented along the tangent of the helical surface and in the 2D plane $(k_x, k_y)$. The rotation of all the spins of the electrons defines the spin tension vortex-skyrmionic state in the BZ. It is noteworthy that the spin rotational movement could be independent of the movement of the electron itself, thereby forming a bundled state attached to the electron. Using the geometrical construction provided in the left picture in Figure 6, one could write, $R_1 = S/2\tan\varphi_1$ , $R_2 = S/2\tan\varphi_2$ , ... $R_{[N/2]} = S/2\tan\varphi_{[N/2]}$, where $R_1, R_2, \ldots R_{[N/2]}$ represent the respective distances of the electrons on the helical surface from the axis of the center-of-mass which is perpendicular to the Cartesian $\boldsymbol{k}ds$-axis. The modulus of the spin vector $|S|$ remains the unchanged physical parameter. It is, therefore, deduced that the value of the product of the radius $R_i$ with the tangent of the angle $\varphi_i$ between $R_i$ and the line that connects the spin vector to the center-of-mass axis remains constant throughout the helical movement of the spin, i.e.

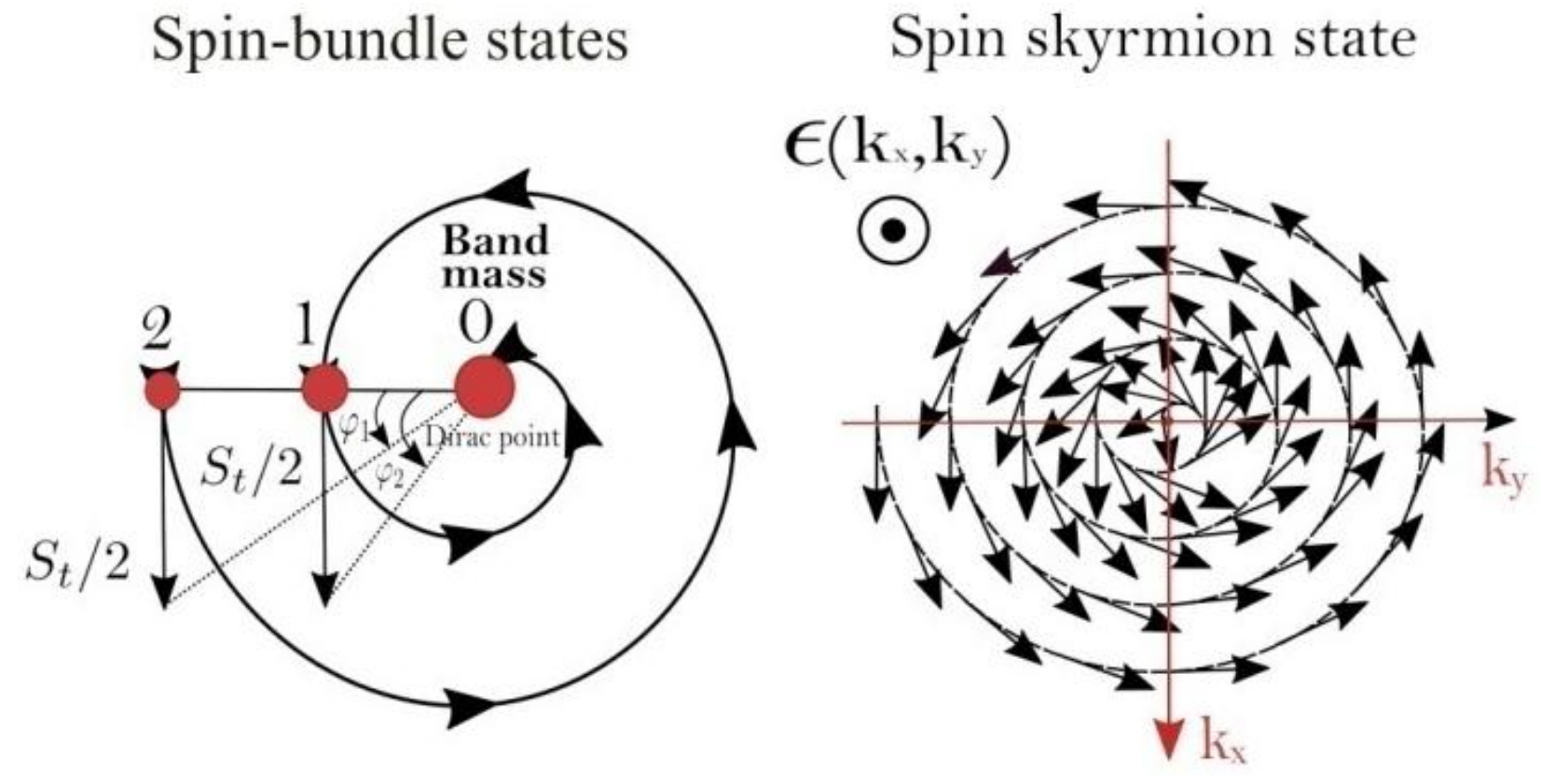

**Figure 6** The spin-bundle state on the helical surface in the conduction band (left-picture) and the spin-skyrmion states on the helical surface (right-picture).

$$|\boldsymbol{R_i}|\tan\varphi_i = const\ , \tag{9}$$

where the index $i$ represents the enumeration of the electrons beginning from the Dirac's point (at which the electrons are attributed a number 0 to denote the reference point) up to the electrons at the Fermi level, denoted as [N/2], where N is the total number of particles and is supposed to be odd in order to have an equal number of electrons in both the branches in the BZ. Therefore, $i = 0, 1, 2, \ldots, \left[\frac{\mathrm{N}}{2}\right] - 1, [\mathrm{N}/2]$ are the electron numbers (Figure 5). The tangent function in Eq. (9) has the periodicity of $\pi$; therefore, a simultaneous relation of type $|\boldsymbol{R_i}|\tan(\varphi_i + \pi) = const$ is also true for any given electron with the number $i$. Therefore, two topologically equivalent points were defined on the opposite branches of the conical surface, and the electrons could occupy these positions with opposite directions of spins (according to the Pauli principle). The product $|\boldsymbol{R_i}|\tan\varphi_i$ may be regarded as the spin tension vector-field of the electron. The term vector is emerges from the fact that the parameter $\boldsymbol{R}_i$ is originally a vector-quantity with a direction same as that of the unit vector perpendicular to the axis at the Dirac's point; it is directed along with the electron's localization position, thereby being parallel to the real axis $|\mathbf{k}|$ds. Indeed, $\varphi_i + 2\pi = \varphi'_{i-1}$, because the angle $\varphi_i$ changes when the electron moves to the neighboring point with the number $k = i - 1$ on the wave vector axis, i.e. $\tan\varphi'_{i-1} \neq \tan\varphi_i$. This is suitably illustrated in the left picture in Figure 6, which depicts the anticlockwise rotation of the electron's spin momentum with variation in angle $\varphi_i$ in the example of two electrons close to the Dirac's point. Therefore, the usual movement of the electron on the band energy branches is accompanied by the simultaneous bundle movement of the spin of the electron on the helical surface. The rotation by $\pi$, of a single spin, leads to the topologically equivalent point situated on the right side branch of the band dispersion and does not describe the periodic movement of the electron's spin-bundle. The topologically equivalent point could lead to the inversion of the spin state if that spin state is not permitted according to the Pauli principle. In the present case, the periodicity of the spin-bundle state would be equal to $4\pi$. When the topologically equivalent state is permitted according to the Pauli principle, the spin-bundle state does not change its direction when rotating around the symmetry axis, and the periodicity of the spin-bundle state attains the value of $2\pi$. Therefore, the rotations by a factor of $2\pi$ and $4\pi$ describe the same electron at different states $\boldsymbol{k}$, closest neighbor and the next-closest neighbor, and represent the correct periodicities of the spin-bundle states according to the spin states at the topologically equivalent points on the helical surface. Such a periodicity of the valley pseudospin, equal to $4\pi$, was observed recently in [12], where it was demonstrated that the pseudospin splitting from the valley-orbit coupling becomes spectrally resolvable in monolayer group-VIB transition-metal dichalcogenides (TMDCs).

## 3. Newton's Laws in Graphene

Next step was to determine the optimal radius $R_v$ for the electron mass-vortex. In order to achieve this, let's abstract for a moment from the usual $\varepsilon(\boldsymbol{k})$-dispersion curves to the mass-dispersion one, i.e., $M(\boldsymbol{k})$-, through the introduction of an Ausgeklugelt $M(\boldsymbol{k})$-dispersion relation. The physical sense of it is well-recognized: for the electrons moving in graphene, one introduces the effective mass of the electron $m^*$. Therefore, the description with the dispersion $M(\boldsymbol{k})$ is well approved and appropriate for that case. In monolayer graphene, $m^*$ is extremely small in comparison to the mass of the free electron, and therefore, the movement of the electrons in monolayer graphene has a quasi-relativistic nature. In the case of BLG, the electron acquires a finite and measurable effective mass, owing to the complicated parabolic 4-band structure in the BLG, and the electron's dressing is less strong. The effective radius of the single-electron mass-vortex confinement would provide an appropriate idea of the localization volume of the electron itself (i.e., a localization space, where the spin-precession is not zero) around the $\boldsymbol{K}$-point in the BZ (and also in general at any point in the $\boldsymbol{k}$-space). It is possible to directly estimate the electron's localization radius close to the electron mass-vortex by considering another probe-electron at point $\boldsymbol{K}_0$ situated on the surface of the sphere with a radius equal to that of the electron band mass-vortex at $\boldsymbol{K}$. In the real space, the probe-electron is separated from the electron at Dirac's point by a distance of $|\boldsymbol{K}-\boldsymbol{K}_0|$ds (Figure 7). The reflection of point $\boldsymbol{K}_0$ with respect of the mass symmetry axis $M(\boldsymbol{k})$ provides the topologically equivalent point on the second branch of the linear dispersion. It should be noted here that the quantity $ds$ is the infinitesimal surface in the Cartesian space $\mathbb{R}_{\mathrm{C}}^{(2)}$ with surface element $ds = dl_x dl_y$ (Figure 3) around the given point $\boldsymbol{K}_0 = (K_{0X}, K_{0Y})$. In general, the surface element $ds$ is different at each point $\boldsymbol{K}_i$ in the reciprocal space; however, here it is supposed that the point $\boldsymbol{K}_0$ is quite close to the Dirac's point, and therefore, the infinitesimal surface $ds$ is same for the points $\boldsymbol{K}$ and $\boldsymbol{K}_0$. The Cartesian space $\mathbb{R}_{\mathrm{C}}^{(2)}$ is the space orthogonal to the reciprocal space $\mathbb{R}_{\mathbf{k}}^{(2)}$, and the direct product of these two spaces generates a 4D manifold, $\mathbb{R}^{(4)} = \mathbb{R}_{\mathrm{C}}^{(2)} \times \mathbb{R}_{\boldsymbol{k}}^{(2)}$, which forms a hypersurface in $S^{(3)}$. Therefore, here a direct product of two parallel spaces is being dealt with– one is the reciprocal $\boldsymbol{k}$-space $\mathbb{R}_{\boldsymbol{k}}^{(2)}$ and the other is the real Cartesian-space $\mathbb{R}_{\mathrm{C}}^{(2)}$. These spaces are depicted in Figure 3, which presents the helical surface of the electron spin rotation. The spaces introduced above are indeed geometrically perpendicular to each other due to the original construction of the 2D reciprocal lattice space, where each of the reciprocal translation vectors $\boldsymbol{G}_i$ (with $i = 1,2$) is perpendicular to the corresponding translation vector in the real space $\boldsymbol{R}_i$ (with $i = 1,2$) and the real space basis in the honeycomb lattice of graphene comprises two translation vectors $\boldsymbol{a_1} = a(3/2, \sqrt{3}/2)$ and $\boldsymbol{a_2} = a(3/2, -\sqrt{3}/2)$. The reciprocal lattice vectors, which form the reference, are represented by the following relations: $\boldsymbol{G}_1 = 2\pi(\boldsymbol{n} \times \boldsymbol{a_2})/|\boldsymbol{a_1} \times \boldsymbol{a_2}|$ and $\boldsymbol{G}_2 = 2\pi(\boldsymbol{a_1} \times \boldsymbol{n})/|\boldsymbol{a_1} \times \boldsymbol{a_2}|$. The symbol "×" in these relations represents the vector product of two vectors $\boldsymbol{G}_1$ and $\boldsymbol{G}_2$. It is clear that $\boldsymbol{G_1} \perp \boldsymbol{a_2}$ and $\boldsymbol{G}_2 \perp \boldsymbol{a_1}$. Furthermore, Newton's gravitational law [48] was written for the probe electron at point $\boldsymbol{K}_0$ and the electron mass-vortex at point $\boldsymbol{K}$ with the band-mass $M_{Band}$. It follows that

$$F_N = G\frac{m^*(\boldsymbol{K}_0)M_{Band}(\boldsymbol{K})}{|\boldsymbol{K}-\boldsymbol{K}_0|^2 ds^2}, \tag{10}$$

where $G$ denotes Newton's gravitation constant, $G \approx 6.674 \times 10^{-11}\, m^3 \cdot kg^{-1} \cdot sec^{-2}$, $m^*(\boldsymbol{K}_0)$ represents the effective mass of the electron at point $\boldsymbol{K}_0$ (this is exactly the effective mass of the electron in graphene), and $M_{Band}(\boldsymbol{K})$ represents the band-mass at point $\boldsymbol{K}$.

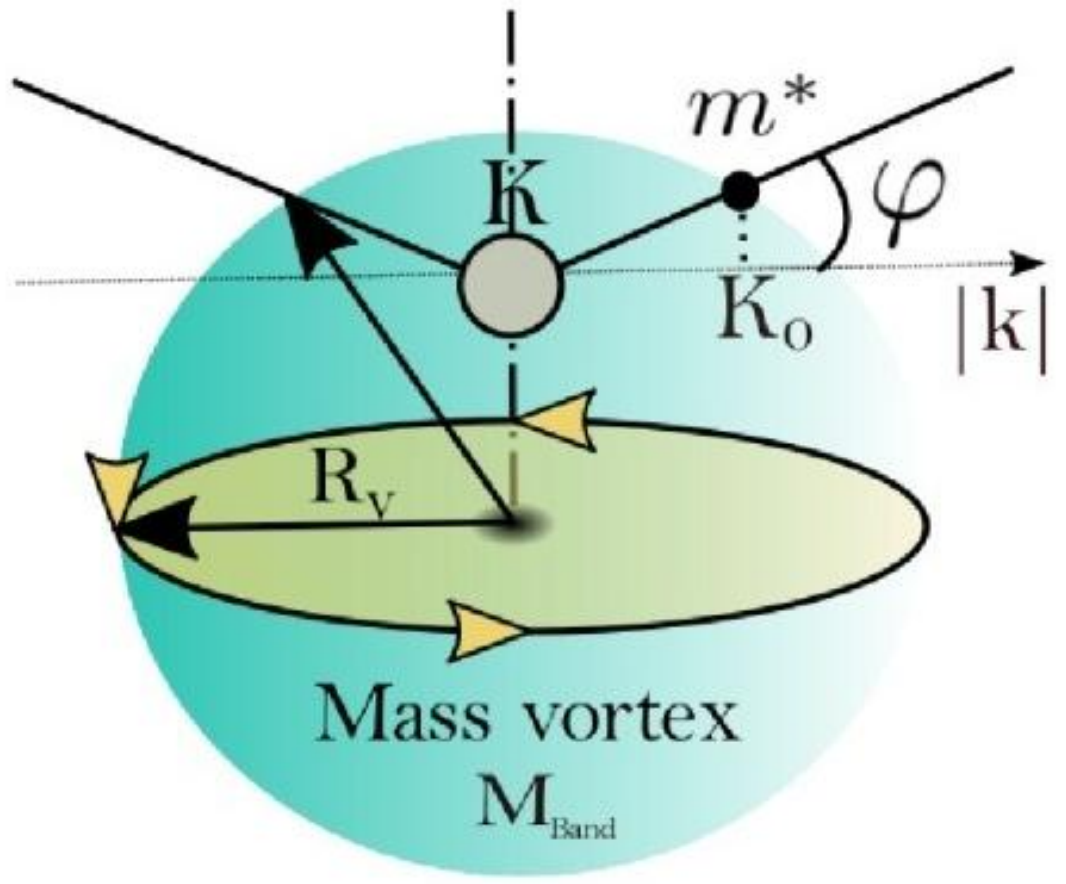

**Figure 7** The electron's mass-vortex at the Dirac's point $\boldsymbol{K}$. The enhancement of the band-mass ($M_{Band}$) at point $\boldsymbol{K}$ is illustrated, while the localization volume of the mass-vortex is depicted in the form of a marine ball with a radius equal to that of the electron's band mass-vortex $R_v$ at the Dirac's point. The probe-electron is placed at point $\boldsymbol{K}_0$ on the sphere and is depicted in the form of a small black ball of mass $m^*$.

Next, an explicit calculation of the electron's mass-vortex localization volume around the Dirac's symmetry point $\boldsymbol{K}$ in graphene was performed. The $M(\boldsymbol{k})$ dispersion scenario introduced above is used here as well. This calculation is achieved by using, the condition to locate two electrons, one at the point $\boldsymbol{K}_0$ and the other at the Dirac's point $\boldsymbol{K}$ (with the corresponding band-mass $M_{Band}$ provided in Section 2.1), which are separated from each other with a distance $R = |\boldsymbol{K} - \boldsymbol{K}_0|ds$, in the 4D space $\mathbb{R}^{(4)} = \mathbb{R}_{\mathrm{C}}^{(2)} \times \mathbb{R}_{\boldsymbol{k}}^{(2)}$. This condition is the subject of Newton's third law [48] in the 4D space $\mathbb{R}^{(4)}$, relating Newton's gravitational force $F_N$ with the spring force $F_{\kappa_G}$ in monolayer graphene, i.e.

$$F_{\kappa_G} = F_N, \tag{11}$$

which leads to the following relation:

$$\kappa_G|\boldsymbol{K}-\boldsymbol{K}_0|ds = G\frac{m^*(\boldsymbol{K}_0)M_{Band}(\boldsymbol{K})}{|\boldsymbol{K}-\boldsymbol{K}_0|^2ds^2}\ , \tag{12}$$

where, $\kappa_G$ is the effective spring constant in monolayer graphene. The condition in Eq. (11) allows for the possibility of the existence of single electronic states close to the band mass-vortex at the Dirac's point. The following expression is obtained for the electron's localization radius: $R_s^3 = |\boldsymbol{K}-\boldsymbol{K}_0|^3ds^3 = Gm^*(\boldsymbol{K}_0)M_{Band}(\boldsymbol{K})/\kappa_G$. As expected, the localization around the Dirac's point depends strongly on the spring constant $\kappa_G$ of the graphene material, and of course, on the band-mass $M_{Band}(\boldsymbol{K})$, which is a non-linear function of the number of electrons $N$ in the conduction band. In case of the BLG, the spring constant is slightly lower, and therefore, the electron localization volume would be larger. It should be recalled that here the simplest case was considered, in which the electron, along with its mechanical momentum and mass-vortex, is localized on the spherical surface of radius $R_v$, which is provided by the electron band mass-vortex at the point $\boldsymbol{K}$ (Figure 7). Then, one would be able to estimate the electron's localization radius $R_s$ in the $M(\boldsymbol{k})$ dispersion scenario using the following expression:

$$R_s = (Gm^*(\boldsymbol{K}_0)M_{Band}(\boldsymbol{K})/\kappa_G)^{1/3} \tag{13}$$

The expression in Eq. (13) provides the localization radius of the individual electron that interacts, via Newton's mass gravitational law, with the band-mass at point $\boldsymbol{K}$ in the BZ. It is possible to estimate the localization radius $R_s$ numerically, by putting values of the physical constants in the expression of $R_s$. The effective mass of electron in graphene is of the order of $m^* = 0.1\times10^{-31}$ kg [44], the spring constant in monolayer graphene varies in the interval $\kappa_G \in (1;5)$ N/m (demonstrated previously in [49]), and it was selected to use $\kappa_G = 2$ N/m here. The value of the band-mass at the Dirac's point may be calculated using the expression in Eq. (3) in Section 2.1, obtaining $M_{Band}(\boldsymbol{K}) = 5.5\times10^{-11}$ kg. Here, simple monolayer graphene with size $S = 1\ cm^2$ , electron concentration $N_{2D} = 1.51\times10^{15}\ m^{-2}$ , and effective electron mass $m^* = 0.012m_e$ (refer to [44] for the experimental results obtained in the Shubnikov-de Haas and Hall effect measurements performed at temperatures up to room temperature) are considered. Finally, the value for the localization radius of the electron is obtained, $R_s = 2.72\times10^{-18}$ m, which is much smaller compared to the radius of the electron, $r_e = 2.81\ \times10^{-15}$ m. It is worth stating that here, the three-dimensional (3D) volume of electron localization is considered, as the precession of the spin of the electron is in 3D. In addition, this is consistent with the formal five-dimensional (5D) space $(\boldsymbol{k}, d\boldsymbol{s}, M(\boldsymbol{k}))$, and the quantity which describes the 5th dimension is a scalar quantity, and for this reason, one has a formal $(4+1)$-dimensional vectorial space. In the language of surface tension dispersion $\epsilon(\boldsymbol{k})$, there is a formal 5D space $(\boldsymbol{k}, d\boldsymbol{s}, \epsilon(\boldsymbol{k}))$, for which the 5th dimension is provided by the surface tension excitation states $\epsilon(\boldsymbol{k})$. The introduction of $M(\boldsymbol{k})$-dispersion is quite useful in the context of Bose-superfluidity in the Bose or Fermi condensate systems, while the consideration of the space $(\boldsymbol{k}, d\boldsymbol{s}, \epsilon(\boldsymbol{k}))$ is significant (see ahead) in obtaining the dispersion in graphene with great precision for the wave vectors in the vicinity of the $\boldsymbol{K}$ -point (the corrections to low-energy linear Dirac's relation). For instance, when considering the formal-5D space $(\boldsymbol{k}, d\boldsymbol{s}, M(\boldsymbol{k}))$, one automatically has the possibility of dealing with the 3D scalar-surface of the dimension of mass for the superfluid vortex (refer to [14]). In the present case, it would be the 3D scalar-mass of the excitonic vortex–anti-vortex paired structure in graphene. The same discussion is valid for the case of BLG as well.

## 4. Dirac's Dispersion Law in Graphene

In this section, the dispersion law in graphene would be obtained by considering the electrons in the first BZ in the conduction band. A similar discussion is also valid for the holes in the valence band. The calculations are begun with the accelerated band-mass dispersion scenario described in Section 2.1 and the electron mass-vortex representation discussed in the previous section. Here, individual electrons are treated as semi-classical, chargeless particles with effective mass $m^*$ and with kinetic energies $\varepsilon(\boldsymbol{k})$, where $m^*$ is the effective mass of the probe-electron. Next, the $4+1$ dimensional space $(\boldsymbol{p}_s, \mathbb{R}_{\mathrm{C}}^{(2)}, \epsilon(\boldsymbol{p}))$ is considered; here, $\epsilon(\boldsymbol{p})$ represents the surface tension states corresponding to a given excited state with momentum $\boldsymbol{p}_s$, which is the impulsion of the particle per unit surface area $dS$ (recall that $dS$ is defined in the real Cartesian space $\mathbb{R}_{\mathrm{C}}^{(2)}$), i.e. $\boldsymbol{p}_s = \boldsymbol{p}/dS$. Using the mass-vortex representation presented in Figure 7 (in Section 3), one may write:

$$\tan\varphi = \frac{\epsilon(\boldsymbol{P_0}) - \epsilon(\boldsymbol{P})}{|\boldsymbol{P}_s - \boldsymbol{P}_{s\mathbf{0}}| v_t}, \tag{14}$$

where $\varphi$ denotes the angle between the $\boldsymbol{p}$ -axis and the surface tension excitation states $\epsilon(\boldsymbol{p})$, and the impulsion $\boldsymbol{P_s}$ corresponds to the impulsion of the electron at the Dirac's point $\boldsymbol{P} = \hbar\boldsymbol{K}$ normalized to the surface area $dS$ of the band mass-vortex at the Dirac's point $\boldsymbol{K}$. The velocity $v_t$ denotes the covariant component of the electron velocity along the impulsion $|\ \boldsymbol{p}\ |$-axis. The relation in Eq. (14) may be rewritten as $\tan\varphi = \frac{\varepsilon(\boldsymbol{P_0}) - \varepsilon(\boldsymbol{P})}{|\boldsymbol{P} - \boldsymbol{P_0}| v_t}$ because of the relation $\epsilon(\boldsymbol{p}) = \frac{\varepsilon(\boldsymbol{p})}{dS}$ between the surface tension states $\epsilon(\boldsymbol{p})$ and the energy dispersion $\varepsilon(\boldsymbol{p})$. On the other hand, one could write an expression which relates the angle $\varphi$ with the distance $R$ in the real space $\mathbb{R}^{(4)}$: $R \in \mathbb{R}^{(4)}$, which is as follows:

$$|\boldsymbol{P} - \boldsymbol{P}_0| dS = R\cos\varphi\ . \tag{15}$$

Next, the trigonometric transformation formula $\tan^2\varphi + 1 = 1/\cos^2\varphi$ is used to combine the relations in Eq. (14) and Eq. (15) in order to eliminate the angle, and one obtains the following expression for the distance in $\mathbb{R}^{(4)}$:

$$R = \frac{dS}{2\hbar v_t}\sqrt{(m^* - M_{Band})^2 v_t^4 + 4\hbar^2 v_t^2 |\mathbf{K} - \boldsymbol{K}_0|^2}\ . \tag{16}$$

It is convenient to verify that $\dim[\mathrm{R}] = [\mathrm{m}]$. The velocity $v_t$ in Eq. (16) is of the order of Fermi velocity $v_F$ in graphene as the probe-electron is supposed to be localized in the area close to the Dirac's point. Furthermore, when projecting the 3-sphere $S^{(3)}$ ($S^{(3)} \in \mathbb{R}^{(4)}$) onto the 2D plane $(|\boldsymbol{k}|, M(\boldsymbol{k}))$, one obtains the circle $S^{(1)} \in \mathbb{R}^{(2)}$, the outermost border of which is denoted as the bundled line $L(S^{(1)})$. Inside the circle $S^{(1)}$ and at the vicinity of the Dirac's point, the effective normalized mass of the electron denoted by $m^*$ tends to the band-mass $M_{Band}$, which is non-linear in $N$: $m^* \to M_{Band}$, while the wave vector $\boldsymbol{K}_0$ tends to $\boldsymbol{K}$. Therefore, inside the circle of radius $R_v$, the electron behaves similar to the band-mass-electron at the Dirac's point, and is, therefore, the same with the band-mass in the space $\mathbb{R}^{(4)}$. Therefore, at the vicinity of the Dirac's point and inside the closed unit circle $S^{(1)}$, one could write the expansion of $R$, as follows:

$$R_1 \approx |\mathbf{K} - \boldsymbol{K}_0| dS \left(1 + \frac{(m^* - M_{Band})^2 v_t^2}{8|\mathbf{K} - \boldsymbol{K}_0|^2 \hbar^2}\right)_{K_0 dS \in S^{(1)}} \approx |\mathbf{K} - \boldsymbol{K}_0| dS \tag{17}$$

In regard to the modulus of the spring force acting on the dispersion branches of graphene, which is related to the mass-vortex enhancement region, close to the Dirac's point, one has $F_\kappa = \kappa_G R_1$, which follows from the classical Hook's law [38]. Moreover, when considering the free-electron just infinitesimally outside of the bundled circle line $L(S^{(1)})$, then $m^* \ll M_{Band}$, and one obtains the second expansion for $R$ in Eq. (16), which is as follows:

$$R_2 \approx \frac{dS}{2\hbar}|m^* - M_{Band}|v_t\left(1 + \frac{2|\mathbf{K}-\boldsymbol{K}_0|^2\hbar^2}{(m^*-M_{Band})^2 v_t^2}\right)_{K_0 dS \notin S^{(1)}} \approx \frac{dS|M_{Band}|v_t}{2\hbar} + \frac{dS\hbar|\mathbf{K}-\boldsymbol{K}_0|^2}{|M_{Band}|v_t} \quad (18)$$

Subsequently, the total normalized energy, i.e., the surface tension, corresponding to the spring force $F_\kappa$ that acts on the unit area of the surface in the mass-vortex location region (at the Dirac point), could be written as:

$$\epsilon(\boldsymbol{k}) = F_\kappa R_2 = \kappa_G R_1 R_2 \approx \kappa_G (dS)^2 \frac{|M_{Band}|v_t}{2\hbar}|\mathbf{K} - \boldsymbol{K}_0| + \kappa_G (dS)^2 \frac{\hbar|\mathbf{K}-\boldsymbol{K}_0|^3}{|M_{Band}|v_t} \quad (19)$$

It is possible to estimate the quantity $\frac{\hbar}{|M_{Band}|v_t}$ appearing in Eq. (18) and Eq. (19), and one obtains $\frac{\hbar}{|M_{Band}|v_t} = 1.41 \times 10^{-29}$ Jsec$^2$/kg m. Therefore, in Eq. (19), the linear Dirac's dispersion relation is obtained, with the correction of the order of $O(|\mathbf{K} - \boldsymbol{K}_0|^3)$ in the wave vector in the reciprocal space. The second term in Eq. (19) could be neglected in comparison to the first linear term. Indeed, the expression in Eq. (19) provides the possibility of estimating the radius of the mass-vortex, $R_v$, at the Dirac's point $\boldsymbol{K}$, for which, the coefficient near $|\mathbf{K} - \boldsymbol{K}_0|$ is equated with the coefficient $\hbar v_F$ obtained from the low-energy expansion around the Dirac's point [50], obtaining the following expression for the surface area of the mass-vortex at $\boldsymbol{K}$:

$$dS = \frac{\hbar\sqrt{2}}{\sqrt{M_{Band}\kappa_G}} = 8.924 \times 10^{-29}\, m^2\,. \quad (20)$$

Then, the corresponding radius of the mass-vortex at point $\boldsymbol{K}$ is $R_v = 5.33 \times 10^{-15}\, m$. The ratio between the radius $R_v$ of the mass-vortex and the electron localization radius $R_s$ is $R_v/R_s = 1.96 \times 10^3$; this order of magnitude is due to the large band-mass $M_{Band}$ formed at the Dirac's point. The value of the radius of the electron mass-vortex is dependent on the band-mass $M_{Band}$ according to the relation in Eq. (20). Finally, it is concluded that at different points $\boldsymbol{k}$ in the BZ, the radius $R_v$ is different. For instance, at the Fermi level $\varepsilon_F$, the radius $R_v$ becomes equal to the radius of the localization of the electron, $R_s$. The schematic representation of the dependence of the mass-vortex radius on the position of the $\boldsymbol{k}$ points in the BZ is provided in Figure 8, below. In addition, this figure illustrates the formation of the excitonic mass-vortex from the individual mass-vortices corresponding to the electron and the hole particles (refer to the inset in Figure 8).

It is indeed possible to estimate the electron band mass-vortex excitation lifetime at the Dirac's point, using the expression of the surface tension excitation states corresponding to electron-hole bonding, i.e., the excitons at the Dirac's point. Supposing the existence of the hole-type mass-vortex at the Dirac's point, is possible to derive the expression for the excitonic binding energy at $\boldsymbol{K}$; for this, let's refer to the quantity defined in Eq. (1) in Section 2 and presented in Figure 1. The rotation momentum $\boldsymbol{\Gamma}$ in the denominator of Eq.(1) is defined by the expression $\boldsymbol{\Gamma} = 2\pi\, \mathbf{R} \times \boldsymbol{v}$. At the Dirac's point, $\boldsymbol{R} = \boldsymbol{K}dS$, where $dS$ denotes the surface area covered by the rotation of the band mass-vortex at point $\boldsymbol{K}$ and its value is provided by Eq. (20). Therefore, the general expression in Eq. (1) for the surface tension $\Delta\epsilon$, corresponding to the electron band mass-vortex at $\boldsymbol{K}$, becomes:

$$\Delta\epsilon = \frac{\kappa_G}{\sqrt{2}|\mathbf{K}||v_F|\Delta\tau}\,. \quad (21)$$

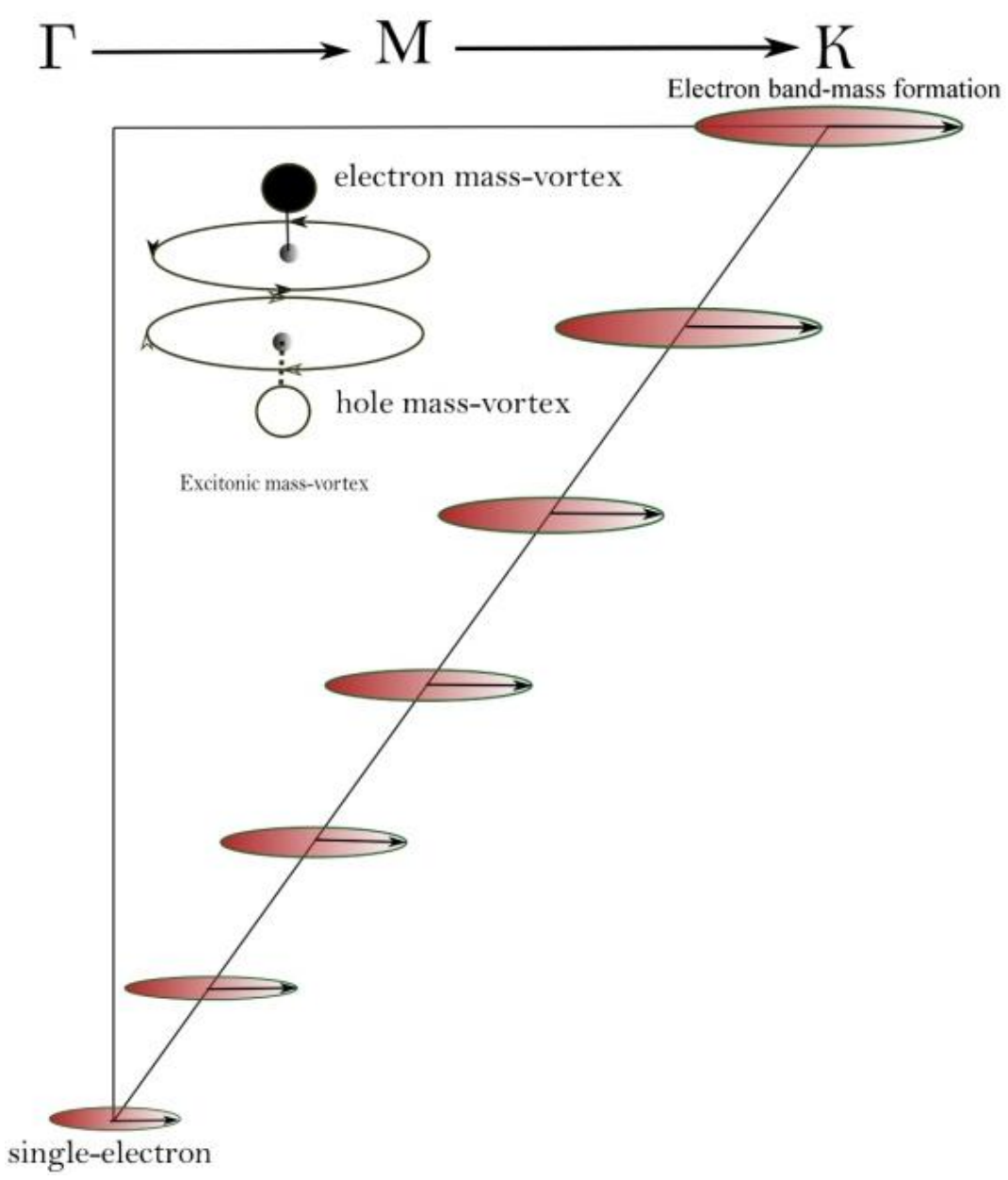

**Figure 8** The electron band mass-vortex formation at the Dirac's point $\boldsymbol{K}$. The circles in light-red depict the consequent formation of the electron mass-vortex along with the high symmetry directions in the BZ ($\Gamma \rightarrow \mathrm{M} \rightarrow \mathrm{K}$) in monolayer graphene. In the inset, the formation of the single excitonic mass-vortex, composed of the electron and hole type mass-vortices with opposite circulation directions, is illustrated.

The hole mass-vortex at the Dirac's point is situated in the conduction band, while the electron mass-vortex is situated in the valence band, which corresponds well to the semi-metallic nature of monolayer graphene (Figure 2 in Section 2). The electron-hole vortex binding energy for a graphene sample with dimension $\Delta S = 10^{-4}\ \mathrm{m}^2$ would be $\Delta\varepsilon_{e-h} = 2\Delta\epsilon\Delta S$. Recently, the first principle calculation results of [51] revealed strong excitonic effects in high-energy photon regime, $E \in (10 - 20)$ eV, in the optical absorption spectra in monolayer graphene. In particular, an extremely weak resonant excitonic peak was observed in the absorption spectrum, and the excitonic binding energy was determined to be $0.43 \times 10^{-19}$ J, in monolayer graphene. Therefore, it is possible to write $2\Delta\epsilon\Delta S = 0.43 \times 10^{-19}$, from which one may obtain the single electron mass-vortex excitation lifetime $\Delta\tau$ provided by Eq. (1). One obtains $\Delta\tau = 3.12 \times 10^{-10}$ s, which is much larger compared to the electron-hole recombination time in graphene ($\sim 10^{-12}$ s) or the mean free path lifetime of the electrons in graphene ($\sim 10^{-14}$ s). The excitation time obtained here, for the electron-hole mass-vortex at $\boldsymbol{K}$, is of the order of the decoupling time of the vertically-stacked pair of spin-vortices with core–core bound states, predicted theoretically in previous studies [52, 53] and observed experimentally in a recent study [54]. Another physical observation concerns the construction of the electron band mass-vortex presented in Figures 1 and 7. Indeed, the ratio between the distance $\Delta x$ [the distance of the electron (or hole) mass-vortices from the particles] and the radius of the mass-vortex $R_v$, i.e. $\Delta x / R_v$, defines the type of polarizability in the system. On the basis of the geometrical structure of the electron representations described above and also the position on the $\boldsymbol{k}$-axis, one obtains two principal polarization states in the electron (or hole) mass-vortex representations. One is related to the case when $\Delta x > R_v$ (or the opposite case, when $\Delta x < R_v$), which is being referred to as the elliptic polarizations of the electron's mass-vortex. The other one is the case when $\Delta x \sim R_v$, which is being referred to as the spherical polarization of the mass vortex. These two polarization states may be detected with the assistance of Bloch sphere rotations under external magnetic field at which the dark excitonic states alter their spin polarizations [55-65] and which causes the enhancement of the optical generation of excitonic valley polarization and macroscopic phase coherence, leading to the excitonic condensate states in monolayer graphene. In such type of measurements, the enhancement of the excitonic states at the Dirac's point could be a direct consequence of the polarized electron and hole mass-vortices, as discussed earlier in the present report. The influence of polarizability of the electron and hole mass-vortices on the formation of excitonic mass-vortices at the Dirac's point is a subject that is out of the scope of the present report and is planned to be considered in the future research to be conducted by our group.

**4. Irreducibility of the Spin-Group**

The dependence of $R_v$ on the $\boldsymbol{k}$ wave vector position has fundamental consequences here, one of which is related to the complex nature of the integrable quasi-complex manifolds, known as the Frölicher-Nijenhuis theorem [66, 67] in the spin-group theory. In the quantum mechanical description of the electron provided in the present report, the direct product $\mathbb{R}_{\mathrm{C}}^{(2)} \times \mathbb{R}_{\boldsymbol{k}}^{(2)}$ is used, leading to the generalized group with the manifold of the cylinder in the 3-sphere, which is irreducible at the Dirac's point, as will be demonstrated ahead. Such calibration of the electron, as a complex isomorphic spin group and as the subgroup of the group of Lies, conserves the principal isomorphic properties associated with the spinorial group, i.e., $\mathrm{Spin}(4)^{\mathrm{R}} \simeq \mathrm{SU}(2) \times \mathrm{SU}(2)$. The addition of the scalar-tensor 2-sphere field comprising the surface-tension-tensor and the mass-tensor (both acting as scalar fields) results in the enhancement of quasi-isomorphism $\mathrm{Spin}(5)^{\mathrm{R}} \simeq \mathrm{Sp}(4)$. In the electron representation provided in the present report, a rotating vectorial field of one-dimension is also added, as a direct product with space $\mathbb{R}^{(4)}$, forming a 5-sphere. The rotating vectorial-field is associated with the electron's mass-vortex rotation around the preferred quantization axis in the spin-group. This further leads to the isomorphism of the type $\mathrm{Spin}(6)^{\mathrm{R}} \simeq \mathrm{SU}(4)$, which is again neither a reducible nor an integrable group of the group of Lies. The spinorial group $\mathrm{Spin}(6)^{\mathrm{R}}$ is a quasi-complex and is not reducible, in the example in the present report, at the $\boldsymbol{K}$-point. This is because of the enlargement of the radius $R_v$ of electron mass-vortex at the Dirac's point (Eq. (13) and Eq. (16) in Section 3), at which an irreducible manifold of two crossing infinite cylinders is formed in the 4D space $\mathbb{R}^{(4)}$ (Figure 9). The formation of such a manifold in 4D space is illustrated in Figure 9; in this figure, there are two perpendicular planes, one of which is the usual Cartesian space $\mathbb{R}_{\mathrm{C}}^{(2)}$ (the directions of the axes of this plane are indicated by the red arrows), while the other space, perpendicular to $\mathbb{R}_{\mathrm{C}}^{(2)}$, is the reciprocal space $\mathbb{R}_{\boldsymbol{k}}^{(2)}$. The manifold of crossing infinite cylinders is formed on the $S^{(3)}$ hypersurface after simultaneous rotations by $\pi$, in the Cartesian and reciprocal spaces, i.e. $S^{(3)} = R_{\pi}(\mathbb{R}_{\mathrm{C}}^{(2)}; (0, R_{0y})) \times R_{\pi}(\mathbb{R}_{\boldsymbol{k}}^{(2)}; (0, k_{0x}))$. Indeed, such rotations $R_{\pi}$ around given principal axes in the 2D spaces $\mathbb{R}_{\mathrm{C}}^{(2)}$ and $\mathbb{R}_{\boldsymbol{k}}^{(2)}$ lead to the formation of crossing perpendicular cylinders in these spaces, which are presented in Figure 9. Therefore,

$$R_{\pi}\left(\mathbb{R}_{\mathrm{C}}^{(2)}; (0, R_{0y})\right) = S_{C}^{(1)} \times I_{R_y},$$

$$R_{\pi}\left(\mathbb{R}_{\boldsymbol{k}}^{(2)}; (0, k_{0x})\right) = S_{\boldsymbol{k}}^{(1)} \times I_{K_x}, \tag{22}$$

where $I_{R_y}$ and $I_{K_x}$ denote the closed unit intervals in the corresponding spaces. Such a procedure of construction of cylindrical manifolds is well-described in [68], where the cylindrical group is discussed in the context of massless particles. The irreducibility of the spinorial group $\mathrm{Spin}(6)^{\mathrm{R}}$ at the Dirac's point provides novel insights into the question of the integrability of the 6-sphere $\mathrm{S}^{(6)}$-manifold (the manifold is not integrable in the present case). Therefore, the 6D generalization is not direct because of the presence of the scalar-rotational $R(S_{m,\varepsilon}^{(2)})$ hypersurface associated with the mass-vortex. In total, the electron represents itself as an irreducible subgroup of the linear group of Lies, i.e., $SL_{el} \in GL(\mathbb{R})$, and $SL_{el} = S_R^{(2)} \times S_{\boldsymbol{k}}^{(2)} \times R(S_{m,\varepsilon}^{(2)})= \mathrm{Spin}(6)^{\mathrm{R}}$ , which is an irreducible subgroup of the structure $\mathrm{S}^{(6)}$ (this is an example demonstrating why the manifold on $\mathrm{S}^{(6)}$ is not integrable, a long-standing problem in mathematics). Another important observation is related to the statement of Frölicher-Nijenhuis [39, 66, 67] regarding the complex nature of integrable quasi-complex manifolds. As demonstrated by the results provided in the present report, the obtained complete 6D manifold is not integrable, and is, therefore, a quasi-complex. The general mechanism underlying this phenomenon is related to the non-vanishing tensor of rotation $R(S_{m,\varepsilon}^{(2)})$, which describes the rotation of the electron's mass-vortex around the quantization axis.

**5. Concluding Remarks**

The present report proposes a novel theoretical approach based on electron and hole mass-vortex representations. This approach considered the band-mass formation at the Dirac's point in graphene, and it was demonstrated that the electron (hole) band-mass is a non-linear function of the number of particles in the given band (the valence or conduction bands, depending on the type of particles considered). It was determined that the electron band-mass was of the order of $M_{Band}(\boldsymbol{K}) = 5.5 \times 10^{-11}\ kg$, which is unusually higher compared to the usual effective mass of the electrons in graphene ($m^{*} \sim 0.012 m_e$). The formation of the band-mass was attributed to the additive nature of the gravitational vector-field in the mass-dispersion representation $\left(|\boldsymbol{k}|, M(\boldsymbol{k})\right)$ in graphene. In addition, the surface tension excitation states in monolayer graphene were considered, and the surface tension of the electron (hole) band mass-vortex at the Dirac's point $\boldsymbol{K}$ was calculated.

Furthermore, the excitonic mass-vortex at the Dirac's point was considered, and its surface tension was evaluated, which permitted the estimation of surface tension excitation lifetime. The electron mass-vortex representation was considered, and relativistic Dirac's dispersion law of the electrons in monolayer graphene was derived, for which the monolayer graphene was considered an elastic medium with the given spring constant and the problem was analyzed classically. The Dirac's energy spectrum in monolayer graphene was derived by considering the classical Newton's laws and Hooke's law, for the single probe-electron, placed at a certain distance [in the elastic medium] from the electron's band-mass, in the reciprocal space. In addition to the usual linear term in the dispersion law, a small correction of the third order was obtained in the wave vector. However, it was demonstrated that this term was negligibly small compared to the linear term. Using Newton's gravitational law, it was determined that the localization radius of the single-electron mass-vortex (which, indeed, provides the real localization radius of the electron) was much smaller compared to the radius of the electron, thereby confirming the existence of strong localization of particles in graphene. Furthermore, similar to the usual linear energy dispersion in graphene obtained in the tight-binding approaches, here also the radius of localization of the electron band mass-vortex at the Dirac's point was determined, and this radius was demonstrated to be much larger compared to the localization radius of the single electron mass-vortex at the Fermi level in graphene. With an interesting intersection of classical physics and quantum mechanical description, the theory developed here provides answers to certain conceptual questions in modern solid-state physics, topology, and spin group theory. In particular, the presence of surface spin tension vectorial field was demonstrated, which is possibly closely related to the surface tension excitation states on the helical surface. It is suggested that the topological phase $\Delta\varphi = \pi$ for the spin-bundle states is protected by the Pauli principle at the position of the topologically equivalent point on the helical surface. The periodicity of the spin-bundle state at the topologically equivalent position was determined to be either $2\pi$ or $4\pi$ , depending on the orientation of the spin. Meanwhile, the periodicity $4\pi$ agrees well with the periodicity of the valley pseudospin, discovered recently in the monolayer transition-metal dichalcogenides [12], which are the ideal materials for the study of Dirac's physics in a solid state. The other fundamental results of the presented theoretical treatment of monolayer graphene were associated with the mathematical aspects of the theory. In particular, there was an unresolved problem in the spin-group theory concerning the formations of manifolds on $\mathrm{S}^{(6)}$ which were suggested as not integrable. According to the results obtained in theory presented here, the electron represents itself as an irreducible subgroup of the linear group of Lies, which follows from the irreducibility of the spinorial group $\mathrm{Spin}(6)^{\mathrm{R}}$ at the $\boldsymbol{K}$-point and is a direct consequence of the band mass-vortex formation at the Dirac's point. Moreover, it was demonstrated theoretically that the spinorial group $\mathrm{Spin}(6)^{\mathrm{R}}$ is a quasi-complex and not reducible, in the example provided in the present report, at the $\boldsymbol{K}$-point. This is because of the enlargement of the localization radius $R_v$ of the electron band mass-vortex at the Dirac's point, at which an irreducible manifold of two crossing infinite cylinders is formed in the 4D space $\mathbb{R}^{(4)}$, which is of the dimension of length. The theory presented here is unique in the sense that it directly reveals the common points of classical Newtonian physics and quantum mechanics using the example of monolayer grapheme.

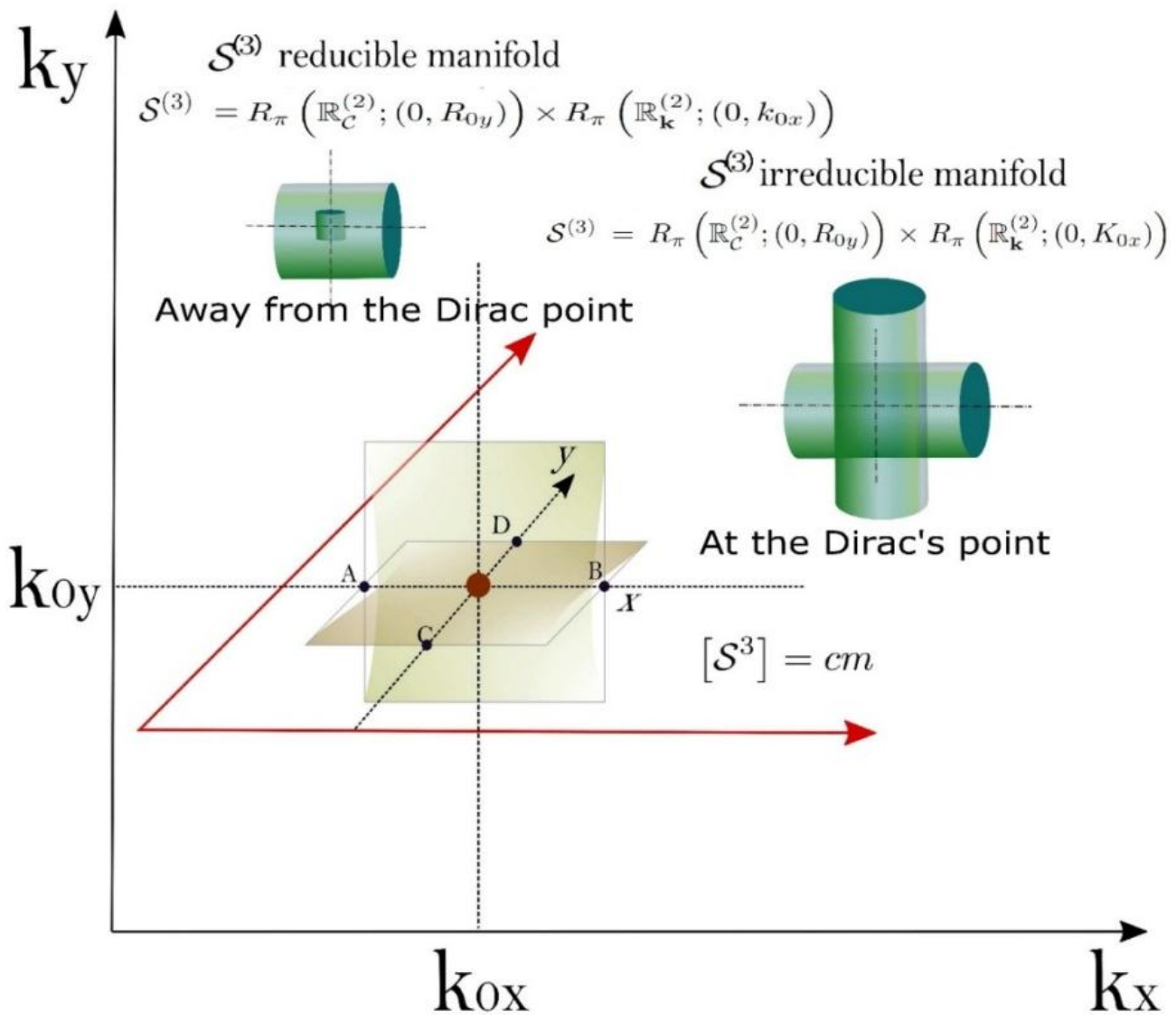

**Figure 9** The formation of crossing perpendicular cylinders after rotations $R_\pi$ around the principal axes in the 2D spaces $\mathbb{R}_{\mathrm{C}}^{(2)}$ and $\mathbb{R}_{\boldsymbol{k}}^{(2)}$.

## Author Contributions

The authors contributed equally to the paper.

## Competing Interests

The authors declare that there are no competing interests regarding this manuscript.